\documentclass[sigplan,nonacm]{acmart}

\usepackage{ragged2e}
\usepackage{graphicx}
\usepackage{xcolor}
\usepackage{url}
\usepackage{soul}
\usepackage{hyperref}
\usepackage{hyperxmp}

\usepackage{tikz}
\usepackage{subfig}
\usepackage{listings}
\usepackage{xcolor}
\usepackage{comment}
\usepackage{multirow}
\usepackage{booktabs}
\usepackage{pgfplots}

\pgfplotsset{compat=1.17}
\usepgfplotslibrary{fillbetween}

\definecolor{codegreen}{rgb}{0,0.6,0}
\definecolor{codegray}{rgb}{0.5,0.5,0.5}
\definecolor{codepurple}{rgb}{0.58,0,0.82}
\definecolor{backcolour}{rgb}{0.95,0.95,0.95}

\lstdefinestyle{mystyle}{
    backgroundcolor=\color{backcolour},   
    commentstyle=\color{codegreen},
    keywordstyle=\color{magenta},
    numberstyle=\tiny\color{codegray},
    stringstyle=\color{codepurple},
    basicstyle=\ttfamily\tiny,
    breakatwhitespace=false,         
    breaklines=true,                 
    captionpos=b,                    
    keepspaces=true,                 
    numbers=left,                    
    numbersep=2pt,                  
    showspaces=false,                
    showstringspaces=false,
    showtabs=false,                  
    tabsize=2
}
\newcommand{\cmmnt}[1]{\ignorespaces}
\newcommand{\sysname}{{\textit{ShadowScope}}}

\usepackage{array}
\newcolumntype{L}[1]{>{\raggedright\let\newline\\\arraybackslash\hspace{0pt}}m{#1}}
\newcolumntype{C}[1]{>{\centering\let\newline\\\arraybackslash\hspace{0pt}}m{#1}}
\newcolumntype{R}[1]{>{\raggedleft\let\newline\\\arraybackslash\hspace{0pt}}m{#1}}

\newcommand*\circled[1]{\tikz[baseline=(char.base)]{
            \node[shape=circle,fill,inner sep=2pt] (char) {\textcolor{white}{#1}};}}

\newcommand{\eg}{\textit{e}.\textit{g}.,~}

\newcommand{\hwname}{\sysname\textsuperscript{$+$}}

\definecolor{yfont}{rgb}{0,0.8,0}
\newcommand{\ghadeer}[1]{\textcolor{blue}{\textbf{Ghadeer: \em #1 }}}
\newcommand{\yicheng}[1]{\textcolor{yfont}{\textbf{Yicheng: \em #1} }}

\newcommand{\nael}[1]{\textcolor{red}{\textbf{Nael: \em #1 }}}
\newcommand{\dmitry}[1]{\textcolor{magenta}{\textbf{Dmitry: \em #1} }}
\newcommand{\dnote}[1]{\textcolor{cyan}{\textbf{David: \em #1 }}}
\AtBeginDocument{%
  }

\acmConference[ASPLOS 2026]{ACM International Conference on Architectural Support for Programming Languages and Operating Systems}{March 22--26,
  2026}{Pittsburgh, PA}

\begin{document}

\title{\sysname: GPU Monitoring and Validation via Composable Side Channel Signals}

\author{Ghadeer Almusaddar}
\authornote{Both authors contributed equally to this research.}
\email{galmusa1@binghamton.edu}
\affiliation{%
  \institution{Binghamton University}
  \city{Binghamton}
  \state{NY}
  \country{USA}
}

\author{Yicheng Zhang}
\authornotemark[1]
\email{yzhan846@ucr.edu}
\affiliation{%
  \institution{University of California, Riverside}
  \city{Riverside}
  \state{CA}
  \country{USA}
}

\author{Saber Ganjisaffar}
\email{sganj003@ucr.edu}
\affiliation{%
  \institution{University of California, Riverside}
  \city{Riverside}
  \state{CA}
  \country{USA}
}

\author{Barry Williams}
\email{bwilli33@binghamton.edu}
\affiliation{%
  \institution{Binghamton University}
  \city{Binghamton}
  \state{NY}
  \country{USA}
}

\author{Yu David Liu}
\email{davidl@binghamton.edu}
\affiliation{%
  \institution{Binghamton University}
  \city{Binghamton}
  \state{NY}
  \country{USA}
}

\author{Dmitry Ponomarev}
\email{dponomar@binghamton.edu}
\affiliation{%
  \institution{Binghamton University}
  \city{Binghamton}
  \state{NY}
  \country{USA}
}

\author{Nael Abu-Ghazaleh}
\email{naelag@ucr.edu}
\affiliation{%
  \institution{University of California, Riverside}
  \city{Riverside}
  \state{CA}
  \country{USA}
}

\begin{abstract}
As modern systems increasingly rely on GPUs for computationally intensive tasks such as machine learning acceleration, ensuring the integrity of GPU computation has become critically important. Recent studies have shown that GPU kernels are vulnerable to both traditional memory-safety issues (\eg buffer overflow attacks) and emerging microarchitectural threats (\eg Rowhammer attacks), many of which manifest as anomalous execution behaviors observable through side-channel signals. However, existing golden model–based validation approaches that rely on such signals are fragile, highly sensitive to interference, and do not scale well across GPU workloads with diverse scheduling behaviors.
To address these challenges, we propose \sysname{}, a monitoring and validation framework that leverages a \textit{composable golden model}. Instead of building a single monolithic reference, \sysname{} decomposes trusted kernel execution into modular, repeatable functions that encode key behavioral features. This composable design captures execution patterns at finer granularity, enabling robust validation that is resilient to noise, workload variation, and interference across GPU workloads. To further reduce reliance on noisy software-only monitoring, we introduce \hwname{}, a hardware-assisted validation mechanism that integrates lightweight on-chip checks into the GPU pipeline. \hwname{} achieves high validation accuracy with an average runtime overhead of just 4.6\%, while incurring minimal hardware and design complexity. Together, these contributions demonstrate that side-channel observability can be systematically repurposed into a practical defense for GPU kernel integrity.
\end{abstract}

\maketitle

\section{Introduction}
\label{sec:intro}






Graphics Processing Units (GPUs) are ubiquitous in modern computing systems, powering everything from mobile devices to large-scale data centers~\cite{gpu_book, gpu_intro}. They accelerate data-intensive and compute-heavy workloads, ranging from multimedia applications to large language models (LLMs) such as ChatGPT~\cite{chat_gpt} and LLaMA~\cite{llama}. To support these demands, GPUs adopt a highly parallel execution model that launches massive numbers of threads~\cite{gpu_book, gpu_simt, cuda_programming_guide}. Given the sensitive nature of many GPU workloads, ensuring the integrity of kernel execution has become increasingly critical.
However, recent studies have shown that GPU kernel execution is vulnerable to both traditional memory-safety issues~\cite{miele2016buffer, buf_of_cgo, gpu_buf_of, mind_control_attack, zhang2024beyond, guo2024gpu, GPU_code_reuse_attack_2025} and emerging microarchitectural attacks~\cite{naghibijouybari2017constructing, zhang2025nvbleed, naghibijouybari2018rendered, zhang2024invalidate, lin2025gpuhammer, frigo2018grand, ravan2025notso}. To protect GPU kernels, researchers have proposed a range of defenses, including both software- and hardware-based approaches~\cite{graviton, jang_hix, lee2022securing, ivanov2023sage}.
However, these solutions often suffer from high performance overhead and are limited in scope, addressing only a subset of attacks.

As an alternative, researchers have explored \textit{golden model-based validation}, which compares a kernel's runtime behavior against a trusted reference derived from known good executions. Prior work has repurposed side channel signals, including performance counters and timing characteristics, and, in controlled laboratory settings, power or EM signals, to build such references~\cite{cathislapd, rad2008power, liu2016code, sehatbakhsh2019emma, nazari2017eddie}. These signals expose execution structure without intrusive software changes. 
Traditional golden model-based validation has so far been evaluated primarily on simple CPUs or small embedded systems in controlled environments, and for programs with invariant execution patterns.
Translating these methods to GPUs is nontrivial.  SIMT execution variability, the presence of many SMs, dynamic scheduling, and other sources of execution variability introduce distorts to traces, making it difficult to capture a single golden model.


\noindent \textbf{Research challenges.} Leveraging side-channel signals for kernel validation on GPUs introduces several challenges due to the complexity of GPU hardware. 
(1) Side-channel signals are fragile and noisy, and interference from concurrent processes can overwhelm the useful signal. Prior studies on small devices even report SNR values below 0.001 for power and EM measurements~\cite{levi2020ask, mangard2004hardware}, and GPU parallelism further complicates matters by overlapping thousands of threads and mixing their behaviors. This makes signal extraction on GPUs especially challenging.
(2) Aligning observations with specific kernel executions is unreliable. Resource contention, kernel scheduling, and driver-level optimizations can shift or stretch execution phases, leading to misaligned traces~\cite{xu2016warped, wang2015simultaneous}. Such misalignment inflates false positives in golden model validation.
(3) Execution patterns vary naturally. Differences across inputs and configurations can change memory access patterns or warp scheduling, which undermines the assumption of stable reference traces. This variability makes it hard to define accurate golden models that generalize across real workloads.

\noindent \textbf{Our approach.} To overcome these challenges, we propose \sysname{}, a GPU monitoring and validation framework based on \textbf{composable golden reference models}. In this design, the full GPU program execution is represented as a sequence of modular segments (\eg segmented along kernel invocations, CPU-GPU memory transfers, or intra-kernel phases). Each segment is validated independently rather than as part of a single monolithic trace. This segmentation reduces the impact of scheduling noise and input variability, making golden models more robust in highly parallel GPU environments.  It also allows for each segment to be configured to allow for some variability in execution.  
\sysname{} builds on two key innovations.
(1) Compositional validation. The golden reference is decomposed into segments defined by composable functions marking kernel boundaries or phases. Each segment captures a repeatable execution pattern and is validated independently, isolating variability and absorbing scheduling noise. An application is deemed valid only if all segments conform to their expected behavior.
(2) Auxiliary instrumentation. The application is instrumented to emit lightweight markers as side-channel signals (more accurately, covert channel signals). These markers indicate segment boundaries and can encode contextual parameters such as kernel and/or input configurations.  When these markers are received by a verifier, they enable it to align the observed traces to improve robustness under concurrency with minimal overhead.  Communicating configuration information as part of the marker can also allow the verifier to dynamically match the appropriate reference golden model to the communicated kernel configuration parameters.  We call this idea composable golden reference models.

We first implemented \sysname{} in software and evaluated it on two NVIDIA GPUs: Tesla V100 and GeForce RTX 4060. Using the NVIDIA CUPTI profiler API~\cite{cupti}, we collect performance data such as timing, instruction counts, and memory usage during kernel execution to construct the golden reference model. We decomposed execution into segments corresponding to kernel invocations and inserted functions that communicate markers through the side channels at the kernel boundaries.
To assess effectiveness, we tested \sysname{} against four representative GPU attacks: buffer overflows~\cite{miele2016buffer, guo2024gpu}, the mind-control attack~\cite{mind_control_attack}, Rowhammer~\cite{mutlu2019rowhammer}, and slowdown/DoS attacks~\cite{taneja2025roguerfm, ravan2025notso}. Across these cases, \sysname{} successfully detected anomalous execution with high accuracy, achieving up to a 100\% true positive rate and as low as 0\% false positives under controlled conditions. Moreover, our evaluation shows that the method is robust to noise and interference from other GPU workloads.

When implementing \sysname{} in software on NVIDIA GPUs, we identified several limitations in the existing Performance Monitoring Units (PMUs) that hinder effective attack detection. These include (1) low sampling rates, which reduce visibility into short-lived kernel behavior, (2) high profiling overhead and restrictions on event groupings imposed by current GPU profiling tools, and (3) Side channels from accessible performance counters: enabling access to performance counters is known to enable side-channel leakage~\cite{naghibijouybari2018rendered}: for this reason, software access to performance counters is often disabled on GPUs.   To address these gaps, we introduce \hwname{}, a lightweight hardware-assisted mechanism for performance monitoring and on-chip kernel validation. 
\hwname{} supports higher sampling rates and isolated profiling events, while eliminating the need for CPU and driver intervention required in software-based validation. These enhancements enable \hwname{} to effectively detect two primary classes of GPU attacks: kernel deviation and mind-control attacks (Section ~\ref{sec:hw_security_eval}). We show that the performance and hardware complexity of \hwname{} are small, making it a practical and efficient solution for securing GPU application execution (Sections ~\ref{sec:hw_performance_eval} and ~\ref{sec:hw_complexity_eval}).

In summary, the paper makes the following contributions:
\begin{itemize}
    \item We present the \textbf{first} defense framework that leverages GPU side-channel signals to validate execution.

    \item We propose a \textit{composable} golden modeling approach that enables robust validation by segmenting execution into independently verifiable units.  These units of execution is demarcated via markers that enables the verifier to synchronize with the execution to tolerate variability in GPU parallel execution.  The markers are communicated through the side-channel.  

    \item We implement \sysname{} in software and show on NVIDIA GPUs that it can detect the presence of four representative attacks with high accuracy.

    \item To improve monitoring quality and lower its overhead, we design a hardware implementation that extends the PMU to implement the monitoring.  
    We show that \hwname{} is able to validate correct execution and identify two major classes of attacks.

\end{itemize}

\section{Background}
\label{sec:background}

In this section, we first review the GPU execution model and then introduce the emerging control flow attacks on the GPU. After that, we describe the GPU Performance Monitoring Units (PMUs) that are accessible to developers and form the fundamental structure of \sysname{}.

\subsection{GPU Execution Security}
Unlike CPU processes, GPU kernels operate using the single instruction, multiple threads (SIMT) model, which enables the execution of thousands of threads in parallel. This improves computation efficiency for workloads such as 3D graphics rendering and deep learning. In this work, we focus specifically on NVIDIA GPUs.

\noindent \textbf{GPU execution.} The execution flow within the GPU begins with the CPU launching a \textit{kernel} on the GPU, specifying the grid and block dimensions. A grid, consisting of multiple blocks, is divided into groups of threads. These blocks are then distributed across the streaming multiprocessors (SMs).
Within each block, threads are organized into warps (32 threads) that execute the same instruction simultaneously. The warp scheduler in each SM selects which warps to execute, ensuring efficient parallel processing. 

\noindent \textbf{{GPU kernel execution integrity.}} Ensuring GPU kernel execution integrity is vital for maintaining the correctness and security of GPU computations. This involves preserving the integrity of both the kernel code and the execution flow. It guarantees that the GPU performs intended operations without malicious interference. 

\noindent \textbf{{Emerging attacks on GPU kernels.}} Recent memory corruption attacks, such as buffer overflow attacks, have been extensively demonstrated on GPUs~\cite{miele2016buffer, gpu_buf_of, mind_control_attack}. These attacks target the crucial part of the GPU execution unit -- the kernel.  For instance, Miele et al.~\cite{miele2016buffer} and Di et al.~\cite{buf_of_cgo} demonstrated heap and stack-based buffer overflow attacks on GPU kernels to corrupt data and manipulate the execution flow. Park et al.~\cite{mind_control_attack} corrupted GPU kernel executions in machine learning models, significantly reducing the accuracy of model predictions.

\subsection{PMU on GPUs}
GPU vendors like NVIDIA and AMD have introduced performance monitoring units similar to those on CPUs to help developers optimize application performance. On NVIDIA GPUs, these performance counters are accessed through the CUDA Profiling Tools Interface (CUPTI)~\cite{cupti}. On AMD GPUs, these counters can be analyzed via the GPU Performance API (GPUPerfAPI)~\cite{amd_GPUPerfAPI}.

\noindent \textbf{{PMU events and metrics.}} NVIDIA GPUs provide two kinds of counters to track the performance of CUDA applications: events and metrics. An event in CUDA CUPTI is a measurable activity that occurs during the execution of a CUDA application~\cite{cupti_doc}. These events gather detailed performance metrics, helping to understand and optimize CUDA application performance. Events and metrics relate to various aspects of GPU architecture resources, including SM, L1/L2 cache, dynamic random access memory (DRAM), GPU interconnects (\eg NVLink and PCIe), and so on. Different events and metrics have varying sampling rates, and some can be profiled together while others cannot. 


\section{System Overview}
\label{sec:kernel_verf}


In this section, we first describe our threat model and then elaborate on the design of our approach, utilizing side channel signals in GPUs to validate untested kernels. 

\subsection{Threat Model and Design Goals}
\label{subsec:threat_model}

We consider a victim and attacker who can launch kernels to run on GPUs. Victim kernels can be part of common applications that can be accelerated on GPUs, such as ML model training and inference~\cite{patwari2022dnn, hu2020deepsniffer, guo2024gpu} or HPC benchmarks~\cite{rodinia1, eastman2017openmm}. The attacker is an unprivileged remote user who can exploit a memory corruption vulnerability to change kernel control flow (e.g., buffer overflow) within the GPU kernels used by the victim application~\cite{miele2016buffer,buf_of_cgo,gpu_buf_of,guo2024gpu}. The attacker may also attempt microarchitectural attacks such as side channel attacks~\cite{naghibijouybari2018rendered, ravan2025notso}, rowhammer attacks~\cite{yao2020deephammer}, DoS attacks~\cite{dosgpu2013}, which originate outside a victim application but try to compromise it by affecting its execution behavior.

\noindent \textbf{Attacker's capabilities.} We assume the attacker operates entirely within the GPU. The attacker can exploit existing memory vulnerabilities to alter the intended execution behavior of the victim’s GPU kernels. 
The attacker does not need to tamper with the program executing on the host CPU. Instead, the attacker can launch malicious kernels on the GPU that run concurrently with the victim’s kernels and attempt microarchitectural attacks to compromise them. We assume that the GPU’s performance monitoring unit (PMU) is trusted and that all commands and data related to the PMU are protected from tampering. In other words, the attacker cannot modify the data collection process or the collected side-channel data, regardless of how many kernels they are able to launch. We monitor the periodic execution of victim kernels using the CUPTI Event API. Notably, CUPTI can only profile kernels that execute within the same context as the profiler~\cite{cuptiEventAPI}.

\noindent \textbf{Design goals.} The goal of the validator of \sysname~is to validate the execution behavior of victim GPU kernels that are vulnerable to existing GPU control flow attacks~\cite{gpu_buf_of, miele2016buffer, mind_control_attack, GPU_code_reuse_attack_2025, guo2024gpu}. We collect side-channel leakages emitted during the execution of untested kernels and compare them against pre-generated golden reference traces. By analyzing deviations between the collected and golden traces, we check whether the execution behavior of the untested kernel has deviated from its execution path. 



\subsection{Overview of \sysname{}}
\begin{figure}[ht]
    \centering
    \includegraphics[width=\linewidth]{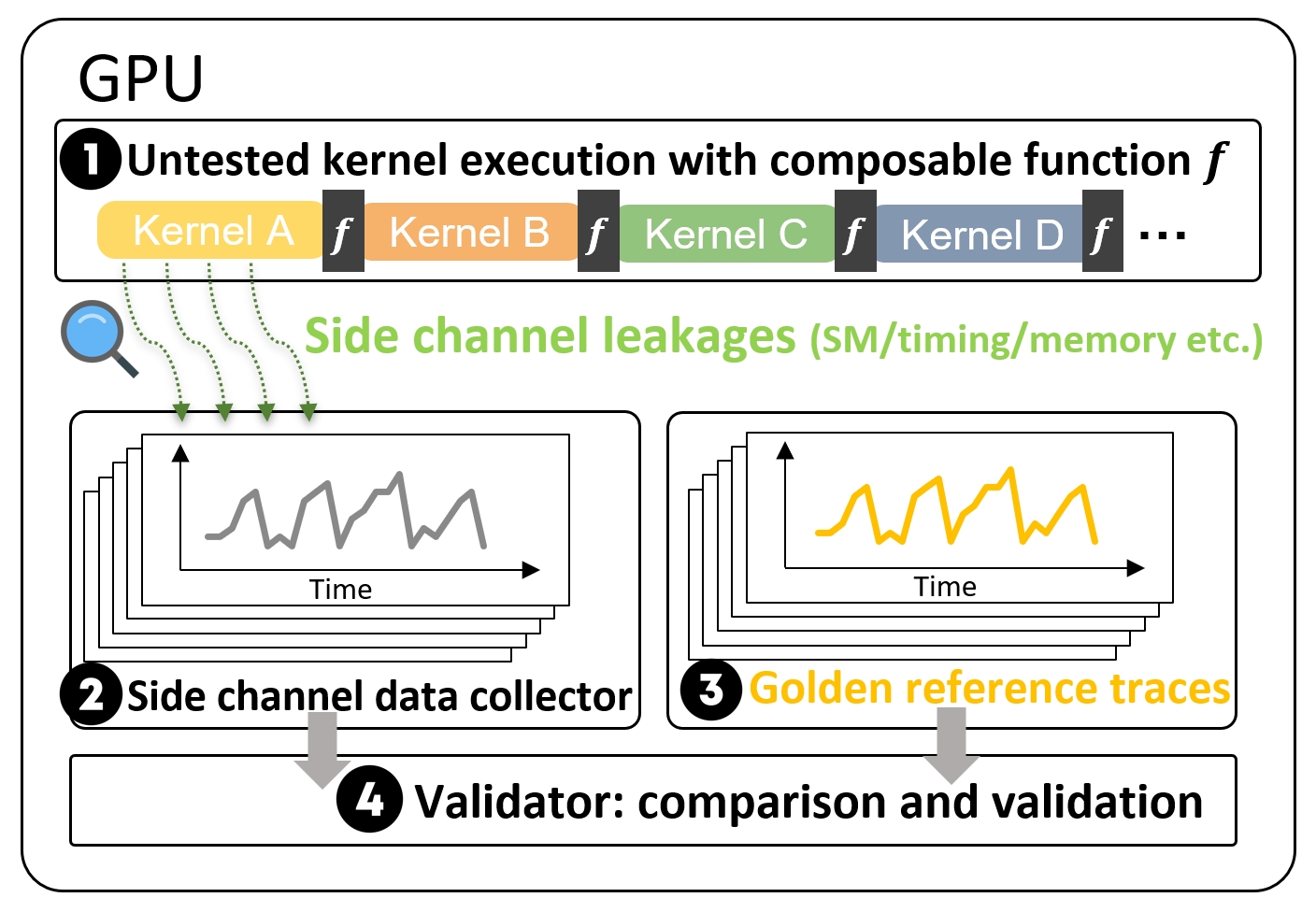}
    \caption{Overview of \sysname{} framework. }
    \label{fig:system_overview}
    
\end{figure}

Fig.~\ref{fig:system_overview} overviews the idea of composable golden models, the validation component of \sysname{} which involves four key components: untested kernels instrumented with composable functions $f$,
a side-channel data collector through PMU, a set of pre-generated golden reference traces of kernels, and a validator. 

We use composable functions to enhance the communication of side-channel information to the validator, supporting more effective golden model validation. This information helps target security-critical kernels more precisely and improves detection accuracy. In addition, it reduces the size of the golden model by focusing only on relevant execution segments. Composable functions can also convey important input parameters, which guide the selection of the appropriate golden model for a given execution context.

The workflow begins when an untested CUDA kernel is executed on the GPU. During execution~\circled{1}, the kernel, instrumented with composable functions, generates side-channel footprints that are captured for validation.
The composable function $f$ is used to mark the start and end boundaries of the untested kernel. Next, the validator’s data collector captures the resulting side-channel signals from the PMU, including the footprints generated by the kernel instrumented with the composable function~\circled{2}. Based on the metadata associated with the execution, the validator then selects the corresponding golden reference trace from a pre-collected dataset~\circled{3}.
Finally, statistical comparison algorithms are applied to compare the collected trace with the corresponding golden trace and determine whether the kernel’s execution integrity has been compromised~\circled{4}. If the execution fails validation, it is flagged as anomalous and prevented from continuing on the GPU.  

\begin{lstlisting}[language=C++, caption=Composable function design., label=listing:comp_func]
__global__ composable_function() {
  int old_val, new_val;
  old_val = atomicCounter;
  new_val = old_val + 1;
  __syncthreads();
  atomicCAS(&atomicCounter, old_val, new_val);
  __syncthreads(); }
void function() {
  int blocks(256,1,1), threads(32,1,1);
  // blocks and threads to execute composable_function
  int compsBlocks(40,1,1), compsThreads(4,1,1);
  // composable function integration
  composable_function<<<cBlocks,cThreads>>>();
  kernel_to_be_validated<<<blocks,threads>>>(arg1, arg2);
  // composable function integration
  composable_function<<<cBlocks,cThreads>>>(); }
\end{lstlisting}

\subsection{Instrumenting GPU Kernels with Composable Functions}
\noindent \textbf{Why insert composable functions?} We insert composable functions $f$  at the beginning and end of each GPU kernel to enhance the robustness of our golden model validation. This design choice is motivated by several reasons:
First, these functions help the validator better align untested traces with their corresponding golden traces. Second, given that GPUs execute hundreds of kernels, precisely identifying the start and end of each kernel enables us to localize potential attacks more effectively.
Furthermore, the composable functions facilitate the transmission of important kernel metadata—such as input size, block size, and grid size—through a covert channel, allowing the validator to select the appropriate golden reference trace for comparison.
Finally, these functions must remain lightweight to avoid significantly altering the kernel’s primary side-channel leakage characteristics.


\noindent \textbf{Design of composable functions on NVIDIA GPUs.} We base these composable functions on atomic operations because they are rarely used in common GPU benchmark kernels and reliably trigger correlated side-channel readings. To identify these functions better, we execute them with a pre-specified number of threads and thread blocks. Composable function design can be adjusted based on the targeted kernel to be distinguishable from kernel execution. For example, we utilize the PMU event \texttt{global\_atom\_cas}, which counts the number of global atomic Compare-And-Swap operations performed on GPU memory~\cite{cupti_doc}. 
This approach enables accurate tracking and monitoring of each kernel's execution boundaries. We explain the integration of composable functions in targeted trusted software in Listing~\ref{listing:comp_func}.

\noindent \textbf{Splitting side-channel traces based on composable functions.}
In addition to events used to validate the kernel, we collect events to detect composable functions. 
This set of events represents the execution behavior of trusted kernels.
For instance, we collect one event group consisting of four events: \texttt{instruction\_executed}, \texttt{global\_store}, \texttt{global\_load}, and \texttt{global\_atomic\_cas}. The first event correlates with the SM level, representing the number of executed instructions in each SM. The \texttt{global\_load} and \texttt{global\_store} events identify the amount of data read from the GPU's global memory by CUDA threads during kernel execution. The last event tracks the number of times an atomic Compare-And-Swap (CAS) operation is executed in global memory. This atomic event helps identify the region of interest based on composable functions within each kernel. 

\begin{figure} [tbh]
    \centering
    \includegraphics[width=0.85\linewidth]{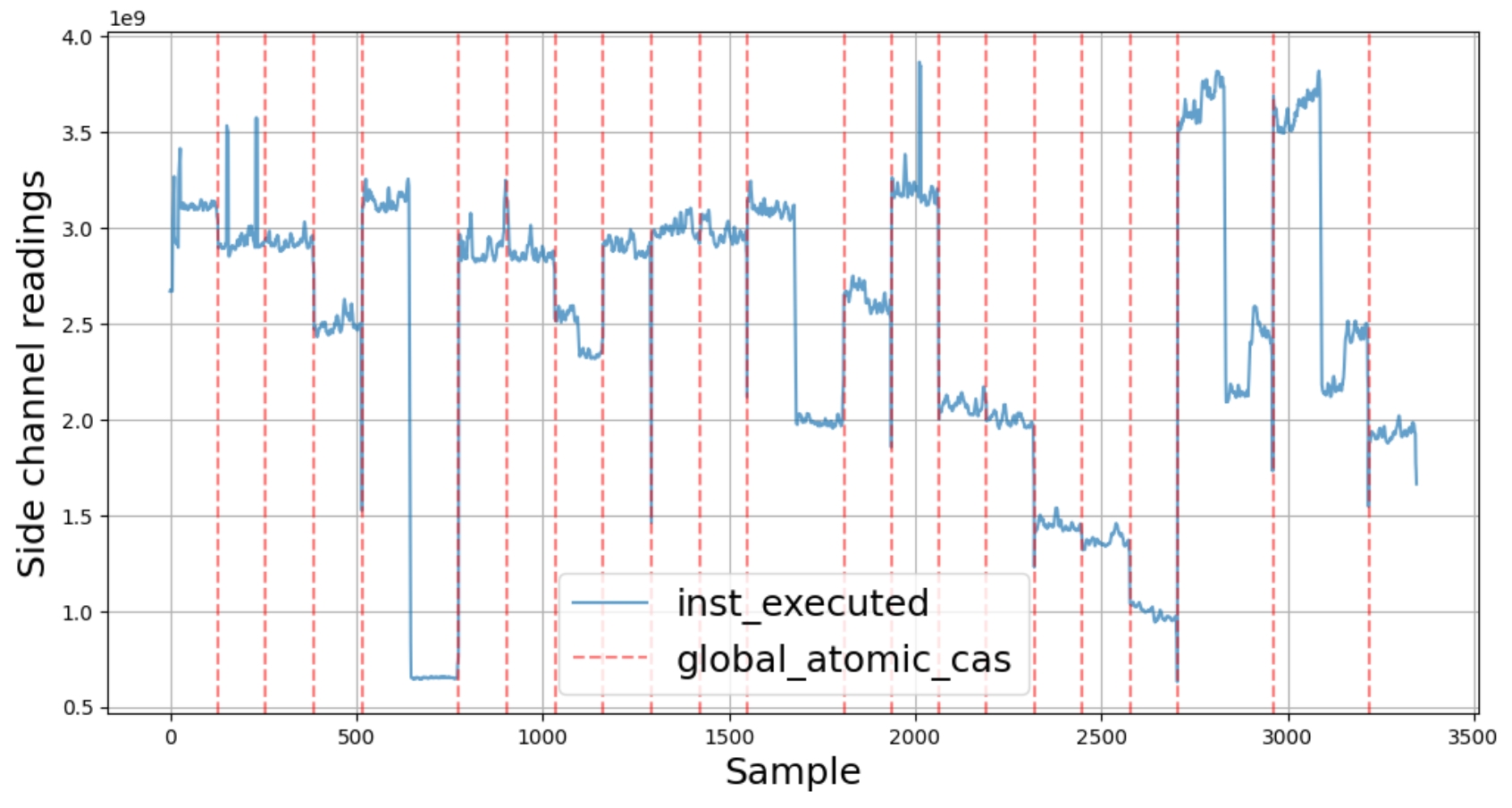}
    \caption{Splitting side-channel traces (AlexNet) based on \texttt{global\_atom\_cas} readings. Each split segment represents a single kernel/layer.}
    \label{fig:split_alexnet}
    \vspace{-5mm}
\end{figure}

As shown in Figure~\ref{fig:split_alexnet}, the readings of \texttt{global\_atom\_cas} help identify the start and end of each layer or kernel execution in AlexNet. For simplicity, we only display the \texttt{instruc\-tion\_executed} readings, which represent the total number of instructions executed across all SMs per sample.

\subsection{Side-channel Data Collector}
\label{subsec:data_collection_events_selection}

\begin{figure}[tbh]

    \centering
    \includegraphics[width=\linewidth]{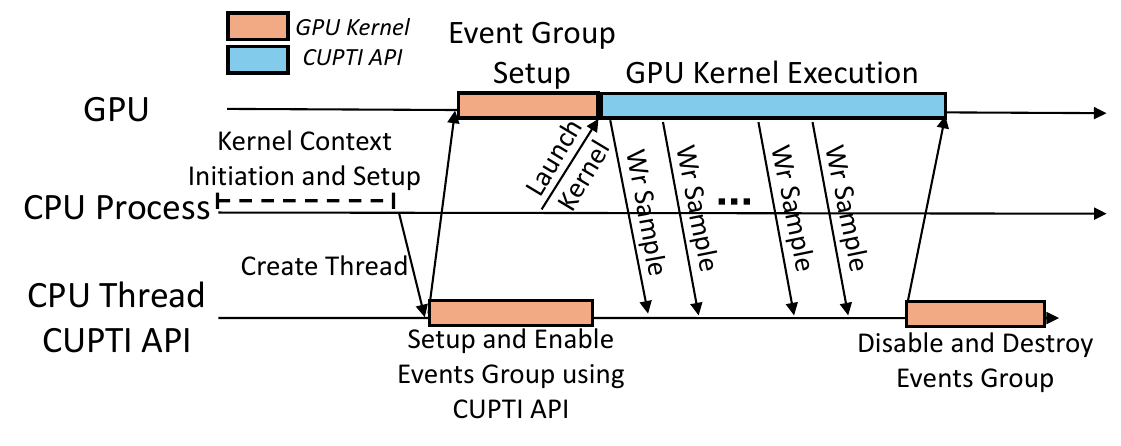}
    \caption{PMU data collection process using CUPTI API.}
    \label{fig:cupti}
    \vspace{-3mm}
\end{figure}

\noindent \textbf{Design of data collector on NVIDIA GPUs.} 
We use the CUPTI API to collect data from performance counters in NVIDIA GPUs. Figure~\ref{fig:cupti} explains the data collection approach using the CUPTI API. Both the targeted kernel and CUPTI profiler need to be in the same context to count the PMU data of the targeted kernel. To count PMU events using the CUPTI profiler, we set data collection mode to continuous using {\fontfamily{qcr}\selectfont cuptiSetEventCollectionMode}. A set of counters needs to be enabled and set to count specific events via  {\fontfamily{qcr}\selectfont cuptiEventGroupCreate}, {\fontfamily{qcr}\selectfont cuptiEventGroupSetAttribute}, and {\fontfamily{qcr}\selectfont cuptiEventGroupEnable}. 
Then, all these events should be added to the \textbf{same events group} for event counts to be read together using {\fontfamily{qcr}\selectfont cuptiEventGroupAddEvent}. An event group is a collection of events that can be counted together. Not all supported events by NVIDIA GPU can be added to the same events group based on their type~\cite{cuptiEventAPI}.

The profiler is set based on the targeted event group and data collection mode. During the execution of the targeted kernel, a CPU thread is used to read events counts per each sample from GPU for collection using {\fontfamily{qcr}\selectfont{cuptiEventGroupReadAllEvents}}. It is important to note that the sampling rate is affected by the type of the event being collected as we will show later.

\subsection{Golden Reference Model of GPU Kernels}

The golden model is a conventional verification technique widely used throughout hardware development life cycles~\cite{he2015model, molina2007functional, kande2022thehuzz}. It verifies whether an IC design satisfies the required specifications. If the design fails to meet these specifications, it must be revised and re-verified until it passes all verification tests.
In this work, we investigate whether golden model verification can be extended to GPU execution flows. Unlike traditional hardware systems, GPUs exhibit fundamentally different architectures and highly parallel execution patterns. To address these differences, we propose a novel golden model specifically tailored for GPU execution, focusing on the kernel, the smallest execution unit on a GPU.

The golden reference model is generated during the secure execution of all trusted kernels and serves as the baseline for validation. Importantly, these golden models only need to be constructed once based on trusted execution behavior.

\subsection{Untested Kernel Validation}


When collecting the execution behavior of an untested kernel using the side-channel data collector based on PMU, the resulting traces are compared against their corresponding golden model. With the help of composable functions embedded within the kernels, we can accurately determine the start and end of each kernel execution. This enables us to extract side-channel readings that precisely correspond to the regions of interest. 

We use cross-correlation as the similarity metric to assess whether the kernel execution flow deviates from the expected behavior. This validation method has been successfully applied in prior work~\cite{sehatbakhsh2019emma}. We choose cross-correlation for two main reasons: (1) it enables pattern matching even when traces are delayed or temporally misaligned, a common occurrence during side-channel data collection via the CUPTI API, because it inherently searches for the optimal lag alignment; and (2) it allows comparisons between traces of unequal lengths, which is critical in GPU environments where concurrent workloads can introduce timing variations and affect the sampling rate.

A signal segment is considered matched if the correlation coefficient exceeds 0.8. To detect attacks, we apply a rejection threshold based on consecutive mismatches. Specifically, \sysname{} flags a kernel as compromised only when four or more rejections occur in succession, allowing it to tolerate up to three consecutive rejections without raising a false alarm.



\section{Software Instantiation of \sysname{} on NVIDIA GPUs}
\label{sec:evaluation_on_real_gpu}

In this section, we first present the software implementation of \sysname{} on NVIDIA GPUs, including side-channel data collection, PMU event selection, and composable golden model generation. We then evaluate its security by implementing four GPU kernel attacks and show that \sysname{} successfully detects all attacks and identifies the compromised kernels. 





\subsection{Events Selection and Grouping} 

Unlike CPU workloads, GPU kernel execution involves multiple Streaming Multiprocessors (SMs), and different scheduling approaches. Some performance events are collected at the level of individual SMs, others capture the aggregate behavior of shared resources across several SMs.
Another challenge with the NVIDIA GPU PMU, particularly when using the CUPTI API, is that not all hardware events can be grouped together~\cite{cupti, cupti_doc}. Based on our findings, events that are collected over the same number of instances (such as per SM or per group of SMs) can be grouped and collected together. For instance, on a Tesla V100 GPU with 80 SMs, the event {\fontfamily{qcr}\selectfont inst\_executed} yields 80 readings, one per SM. In contrast, the event {\fontfamily{qcr}\selectfont fb\_subp0\_read\_sectors} produces only 16 readings per sample, as each reading represents a group of 5 SMs.
The NVIDIA Volta microarchitecture supports profiling 82 events. We categorize these events in Table~\ref{tab:volta_events}.

\begin{table*}[h]
    \small
    \centering
    \caption{CUPTI Events in NVIDIA Volta.}
    \label{tab:volta_events}
    \begin{tabular}{||p{0.15\textwidth}|p{0.8\textwidth}||}
    \hline
    \centering\textbf{Category} & \textbf{CUPTI Events} \\
    \hline
    \hline
    \centering\textbf{SM}& {\RaggedLeft active/elapsed\_cycles\_pm/sys/warps/sm,  inst\_executed/issued1/issued0,  thread\_inst\_executed, sm\_cta/warps\_launched}\\
    \hline
    \centering\textbf{Memory} & {\RaggedLeft fb\_subp0/subp1\_read/write\_sectors}\\
    \hline
    \centering\textbf{FP Unit} & {\RaggedLeft inst\_executed\_fma/fp16\_pipe\_s0/s1/s2/s3} \\
    \hline
    \centering\textbf{L2 Cache}& {\RaggedLeft l2\_subp0/subp1\_read/write/total\_sector\_misses/queries, l2\_subp0/subp1\_write\_sysmem\_sector\_queries}\\
    \hline
    \centering\textbf{Tensor Cores}& {\RaggedLeft tensor\_pipe\_active\_cycles\_s0/s1/s2/s3}\\
    \hline
    \centering\textbf{Trigger Unit}& {\RaggedLeft prof\_trigger\_00/01/02/03/04/05/06/07}\\
    \hline
    \centering\textbf{Atomic Ops}& {\RaggedLeft atom\_count, global\_atom\_cas, shared\_atom, shared\_atom\_cas}\\
    \hline
    \centering\textbf{Global Memory}& {\RaggedLeft global\_load/store}\\
    \hline
    \centering\textbf{Local Memory}& {\RaggedLeft local\_load/store}\\
    \hline
    \centering\textbf{Shared Memory}& {\RaggedLeft shared\_load/store, shared\_ld/st\_bank\_conflict, shared\_ld/st\_transactions} \\
    \hline
    \centering\textbf{Texture}& {\RaggedLeft l2\_subp0/subp1\_read/write\_tex\_hit\_sectors, l2\_subp0/subp1\_read/write\_tex\_sector\_queries}\\
    \hline
    \centering\textbf{PCIe}& {\RaggedLeft pcie\_rx/tx\_active\_pulse}\\
    \hline
    \centering\textbf{misc}& {\RaggedLeft generic\_load/store} \\
    \hline
    \end{tabular}

\end{table*}

\noindent \textbf{{Targeted GPU kernels.}}
To evaluate our golden model approach \sysname{}, we targeted benchmarks from different suites. These benchmarks are written in the CUDA programming language. We categorize 18 benchmarks from Rodinia~\cite{rodinia1}, CUDA-SDK~\cite{cudaSamples}, GraphBig~\cite{nai2015graphbig} and four famous DNN models, summarized in Table~\ref{tab:benchmarks}. 

\begin{table}[h!]
\centering
\small
\caption{Evaluated benchmarks.}
\resizebox{\linewidth}{!}{
\begin{tabular}{||l|p{5cm}||}
\hline
\textbf{Suite} & \textbf{Benchmarks} \\
\hline \hline
\textbf{Rodinia} & \RaggedRight{gaussian, heartwall, huffman, lud, myocyte, particlefilter, srad}\\
\hline
\textbf{CUDA SDK} & \RaggedRight{matrixMul, vectorAdd, convolutionSeperable, histogram, sortingNetworks, fp16ScalarProduct} \\
\hline
\textbf{GraphBig} & \RaggedRight{BetweennessCentr} \\
\hline
\textbf{DNN Models} &  \RaggedRight{CifarNet, AlexNet, SqueezeNet, ResNet-50}\\
\hline
\end{tabular}}
\label{tab:benchmarks}
\end{table}

\subsection{Targeted Attacks}
In this work, we evaluate \sysname{} against four types of attacks targeting GPU kernel execution and demonstrate its effectiveness in identifying abnormal or modified kernel execution flows. These attacks fall into the following categories. 


\noindent \textbf{{Attack 1: buffer overflow attack.}} 
Buffer overflow attacks occur when an attacker overwrites the call stack's return address, redirecting the original program execution to malicious code. This type of attack was first introduced in prior work~\cite{miele2016buffer, guo2024gpu}. It involves several steps. (1) The attacker exploits stack-based buffer overflow vulnerabilities in the CUDA kernel by providing input that exceeds the size of a fixed buffer, causing it to overwrite adjacent memory on the stack, such as function pointers. (2) The attacker crafts the input to overwrite these pointers with addresses of malicious code. (3) The payload is triggered by invoking the vulnerable kernel function. As a result, when the kernel executes, it uses the overwritten function pointers, which now point to the attacker's code. This causes the malicious payload to run instead of the intended function, allowing the attacker to take control of the CUDA application's execution flow.

\noindent \textbf{{Attack 2: Mind control attack.}} The mind control attacks~\cite{mind_control_attack} involve exploiting GPU memory vulnerabilities to undermine deep learning model performance. The attack process includes several key steps. (1) Setup: Attackers gain arbitrary code execution on the GPU by exploiting memory vulnerabilities in GPU kernels. This often involves hijacking the control flow of the GPU kernel using techniques like buffer overflow to overwrite function pointers. (2) Skipping GPU kernel execution: Attackers overwrite the identified GPU kernels, effectively converting them into \texttt{no-ops}. This manipulation degrades the deep learning model's accuracy, causing predictions to become no different from random guessing.

\noindent \textbf{Attack 3: Rowhammer attack.} Rowhammer is a hardware-level attack that rapidly accesses DRAM rows to induce bit flips in adjacent rows~\cite{mutlu2019rowhammer}. These bit flips can corrupt data or be exploited to bypass security mechanisms. Prior work~\cite{yao2020deephammer, lin2025gpuhammer} has shown that such flips in DNN model weights can significantly degrade model performance. In this work, we aim to protect GPU memory against Rowhammer attacks. We launch such attacks on the GPU in two main steps: (1) evicting the L2 cache using the \texttt{discard} instruction~\cite{PTX} (available only on NVIDIA's Ampere and Ada architectures), and (2) repeatedly accessing the same DRAM bank on the GPU thousands of times to trigger potential bit flips.

\noindent \textbf{{Attack 4: DoS/Slow-down attack.}} Modern DRAM chips (e.g., DDR5 and GDDR5) have introduced a Refresh Management (RFM) interface to help mitigate Rowhammer attacks. However, recent work~\cite{taneja2025roguerfm, ravan2025notso, nazaraliyev2025practical} has shown that RFM features can be exploited by attackers to intentionally trigger refresh activity and block DRAM banks, thereby slowing down co-located applications. These attacks are both effective, causing up to a 4.8× slowdown, and stealthy, as the attacker can activate only a single address in different RFM sub-banks to evade detection. Similar to traditional Rowhammer attacks, we simulate this attack on the GPU in two main steps: (1) flushing the L2 cache, and (2) randomly accessing DRAM addresses in RFM sub-banks to trigger slowdowns in victim applications.

\subsection{Evaluation Results}
\noindent \textbf{{Experiment setup and data collection.}} We conduct Attacks 1 and 2 on an NVIDIA Tesla V100 GPU with driver version 535.183.01 and CUDA version 12.2. Attacks 3 and 4 are performed on a GeForce RTX 4060 with Samsung GDDR6 memory, using driver version 545.29.06 and CUDA version 12.3.  To evaluate our defense method, we collect three datasets for each benchmark: golden, normal, and attack. Each dataset contains 100 traces. The golden and normal datasets are collected under benign conditions using only GPU benchmarks. In contrast, the attack dataset is recorded during active attack execution. We use the golden dataset to extract reference traces. Then, we evaluate detection performance using the normal and attack datasets. The golden model comparison algorithm is implemented in Python. True Positive Rate (TPR) and False Positive Rate (FPR) are used as evaluation metrics.


\subsubsection{Evaluation of Buffer Overflow Attack (Attack 1)}
\label{subsec:attack1}
In this attack, the adversary exploits a buffer overrun to overwrite function pointers and redirect execution flow. Each call to a targeted function can be hijacked to invoke attacker-controlled code. To monitor the attack, we utilize four CUPTI events per SM: {\fontfamily{qcr}\selectfont global\_load}, {\fontfamily{qcr}\selectfont global\_store}, {\fontfamily{qcr}\selectfont inst\_e\-xecuted}, and {\fontfamily{qcr}\selectfont global\_atom\_cas}. Since the Tesla V100 has 80 SMs, each reading yields 320 samples. We use {\fontfamily{qcr}\selectfont glo\-bal\_atom\_cas} as a marker to identify the start and end of each kernel. 

\begin{figure}[ht]
    \centering
    \includegraphics[width=\linewidth]{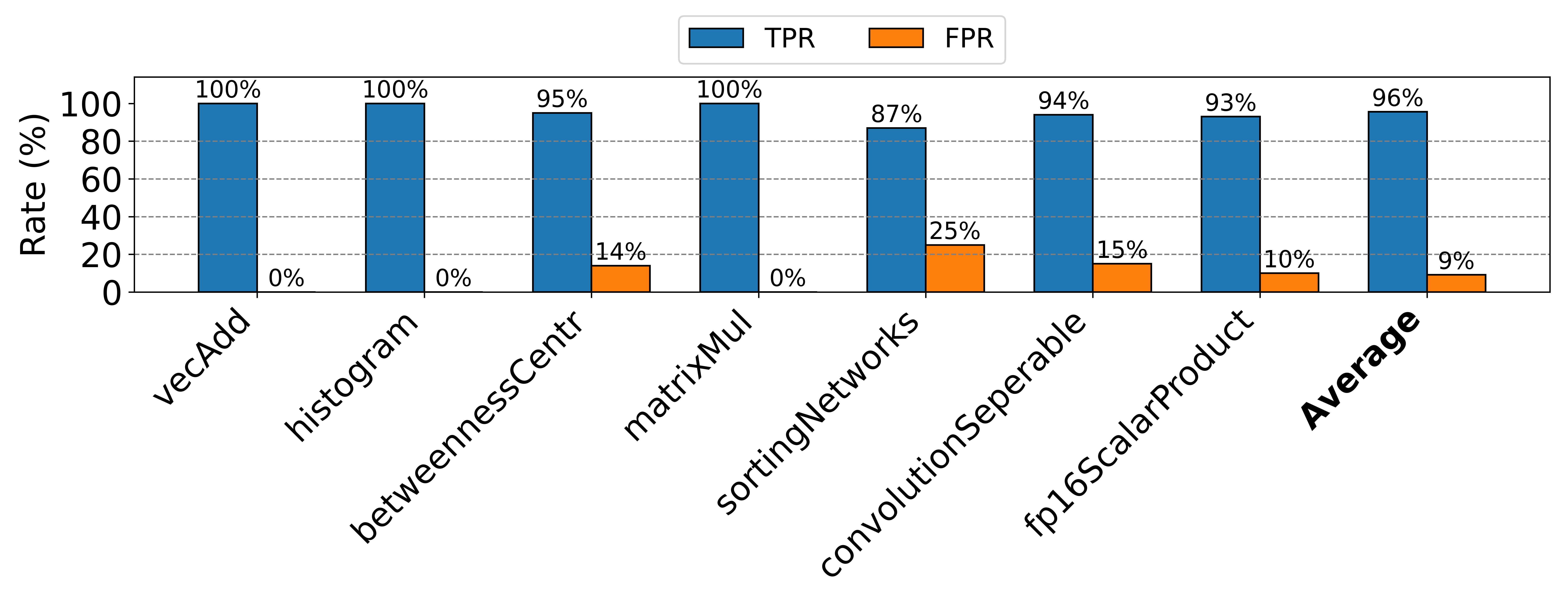}
    \caption{Performance of \sysname{} in monitoring buffer overflow attacks.}
    \label{fig:attack_1_results}
    \vspace{-2mm}
\end{figure}

Figure~\ref{fig:attack_1_results} presents the performance of \sysname{} in identifying buffer overflow attacks across selected benchmarks from the CUDA SDK and GraphBig suites. As shown, \sysname{} consistently achieves high TPR, with 3 out of 7 benchmarks reaching 100\%, and the lowest still maintaining a strong 87\% (\textit{sortingNetworks}). The False Positive Rate (FPR) remains low overall, with 4 benchmarks recording 0\%, and the highest FPR observed being 25\% for \textit{sortingNetworks}. The average TPR and FPR are 96\% and 9\%, respectively.  The relatively high FPR observed in some benchmarks is caused by the limited sampling rate of the GPU’s PMU. In some fast-executing kernels, such as those from \textit{sortingNetworks} or \textit{convolutionSeparable}, the PMU collects fewer than 20 readings per kernel. This low resolution makes it difficult to compare traces against the golden reference accurately. As a result, false positives increase. In Section~\ref{sec:gpu_centric_shadowscope}, we propose a new GPU PMU design to address this limitation.

\subsubsection{Evaluation of Mind Control Attack (Attack 2)}

In this attack, critical DNN layers or GPU kernels are either skipped or replaced with no-op operations. To detect such behavior, we use the same four CUPTI events described in Section~\ref{subsec:attack1} to monitor GPU kernel executions. 

\begin{figure}[ht]
    \centering
    \includegraphics[width=\linewidth]{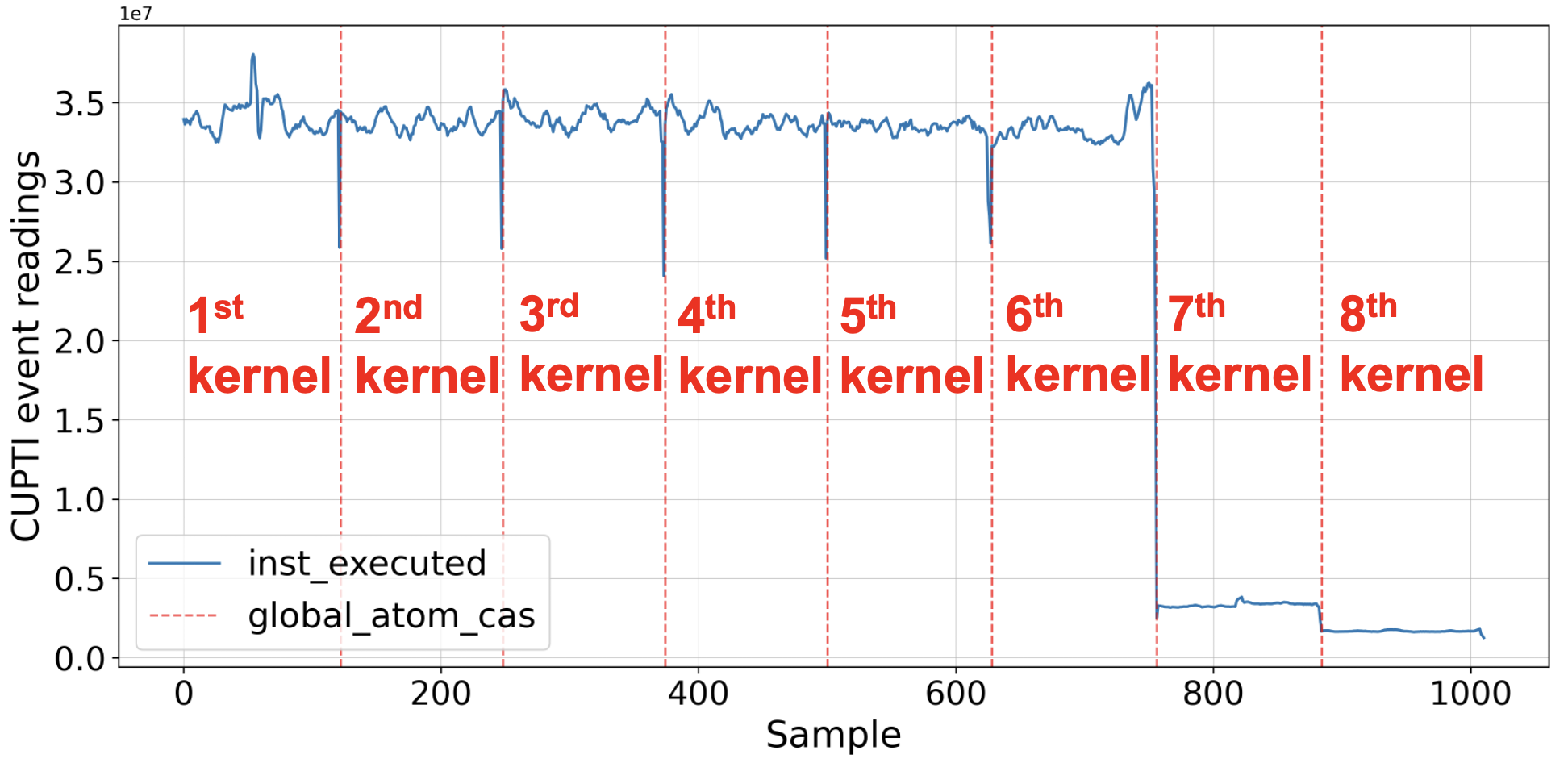}
    \caption{Side-channel signal of normal \textit{CifarNet} execution. }
    \label{fig:cifar_normal}
\end{figure}

\begin{figure}[ht]
    \centering
    \includegraphics[width=\linewidth]{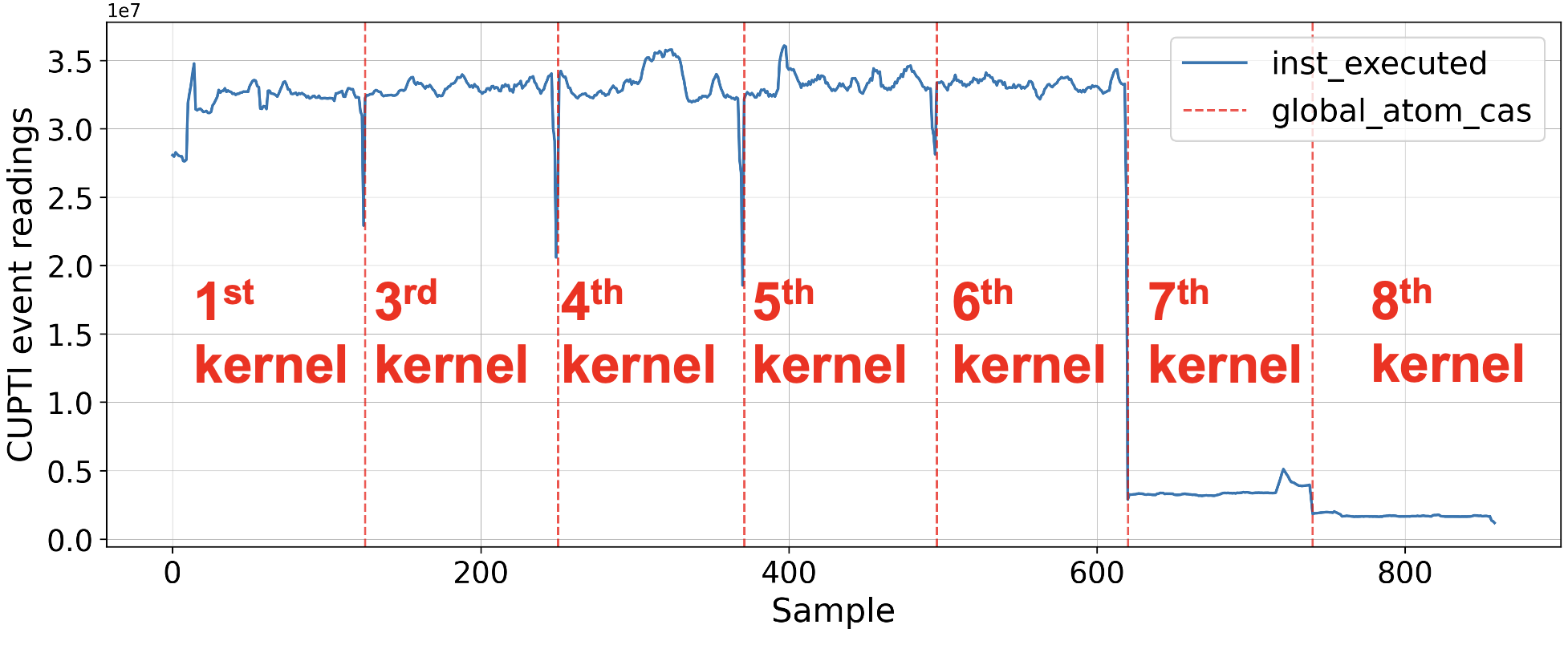}
    \caption{Side-channel signal of \textit{CifarNet} under attack. \textbf{The second kernel is skipped.}}
    \label{fig:cifar_attack}
\end{figure}

Figures~\ref{fig:cifar_normal} and~\ref{fig:cifar_attack} compare the CUPTI event traces of normal and attack executions for CifarNet. In the normal trace (Figure~\ref{fig:cifar_normal}), eight distinct kernel executions are observed. Each is clearly separated by spikes in the {\fontfamily{qcr}\selectfont global\_atom\_cas} event and exhibits consistent {\fontfamily{qcr}\selectfont inst\_executed} activity. In contrast, the attack trace in Figure~\ref{fig:cifar_attack} deviates from this pattern. Although the {\fontfamily{qcr}\selectfont global\_atom\_cas} event still marks the kernel boundaries, the second kernel is missing. This layer-skipping attack shifts the execution sequence and introduces irregular fluctuations in {\fontfamily{qcr}\selectfont inst\_executed} values as well. \sysname{} detects such anomalies by capturing these structural and behavioral inconsistencies.

\begin{figure}[ht]
    \centering
    \includegraphics[width=\linewidth]{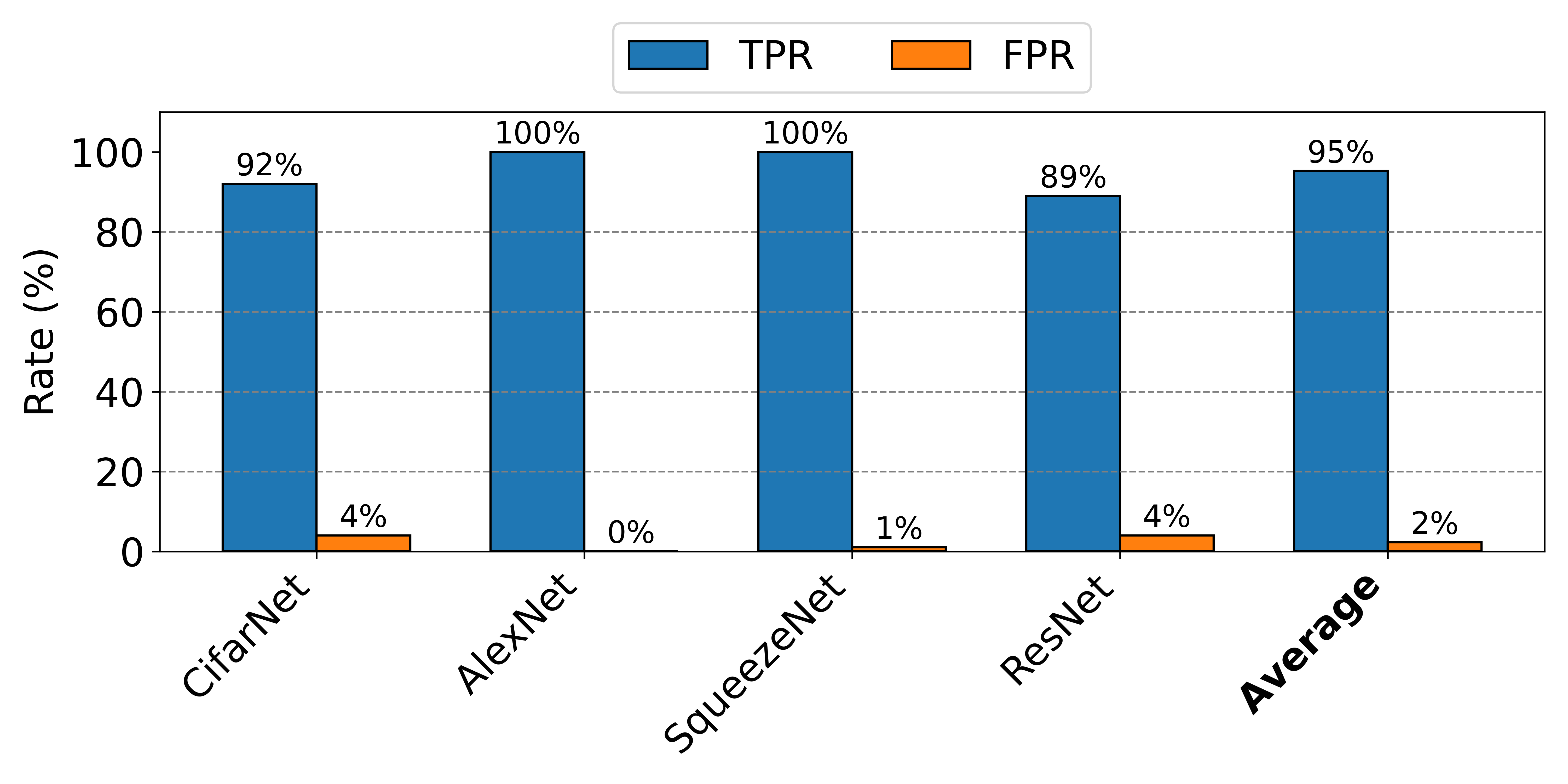}
    \caption{Performance of \sysname{} in monitoring mind control attacks.}
    \label{fig:attack_2_results}
    \vspace{-5mm}
\end{figure}

Figure~\ref{fig:attack_2_results} shows the detection performance of \sysname{} across four representative DNN architectures: CifarNet, AlexNet, SqueezeNet, and ResNet-50. \sysname{} achieves perfect detection (100\% TPR) on both AlexNet and SqueezeNet, with zero or near-zero false positives (less than 1\%). Detection on CifarNet and ResNet remains strong, reaching 92\% and 89\% TPR, respectively. However, both exhibit slightly higher FPRs of 4\%. Overall, the system delivers consistent results, with an average TPR of 95\% and an average FPR of only 2\%.

\subsubsection{Evaluation of Rowhammer Attack (Attack 3)}
\label{subsec:attack3}
In this attack, we demonstrate how \sysname{} can detect abnormal memory behavior, such as Rowhammer attacks, within the GPU’s GDDR memory. To monitor memory activity, we use four CUPTI events: {\fontfamily{qcr}\selectfont \small fb\_subp0\_read\_sectors}, 
{\fontfamily{qcr}\selectfont \small fb\_subp1\_read\_sectors}, 
{\fontfamily{qcr}\selectfont \small l2\_subp0\_total\_read\_se\-ctor\_queries}, and {\fontfamily{qcr}\selectfont \small l2\_subp0\_total\_write\_sector\_qu\-eries}.
These events help track GPU memory usage patterns at both the DRAM and L2 cache levels. Rowhammer attacks rely on frequent row access and L2 cache flushing. As a result, they produce distinct memory access patterns that \sysname{} can effectively capture. 

\begin{figure}[t]
    \centering
    \includegraphics[width=\linewidth]{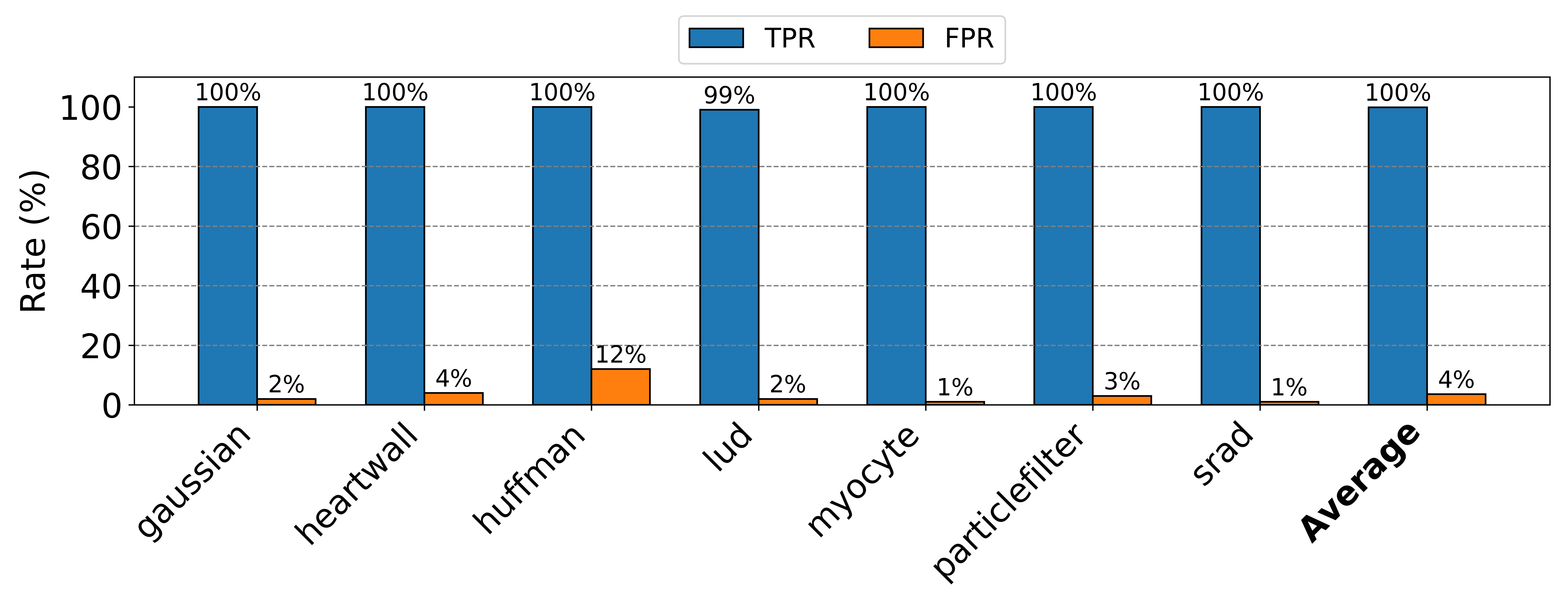}
    \caption{Performance of \sysname{} in monitoring rowhammer attacks.}
    \label{fig:attack_3_results}
    \vspace{-2mm}
\end{figure}

Figure~\ref{fig:attack_3_results} shows the detection performance of \sysname{} across seven benchmarks from the Rodinia suite. \sysname{} achieves perfect detection (100\% TPR) on six out of seven benchmarks, with \textit{lud} slightly lower at 99\% TPR. Most workloads exhibit low false positive rates, ranging from 1\% to 4\%, except for \textit{huffman}, which records a higher FPR of 12\%. We observe that when the sampling rate falls below 50 samples per kernel, detection accuracy decreases, which leads to more false positives in \textit{huffman}.
Overall, \sysname{} demonstrates strong and consistent performance across diverse workloads, achieving an average TPR of 100\% and an average FPR of only 4\%.

\subsubsection{Evaluation of DoS/Slow-down Attack (Attack 4)}

In this attack, we use the same four memory-related CUPTI events described in Section~\ref{subsec:attack3} to detect abnormal GDDR memory usage by the adversary. These events help monitor low-level memory access behavior. Recent denial-of-service (DoS) or slowdown attacks~\cite{taneja2025roguerfm, ravan2025notso} on DDR memory exploit spoofed RFM to block access to DRAM banks. This behavior can severely degrade the performance of co-located applications. To succeed, the attack must both flush the L2 cache and randomly access DRAM addresses within RFM sub-banks. 

\begin{figure}[t]
    \centering
    \includegraphics[width=\linewidth]{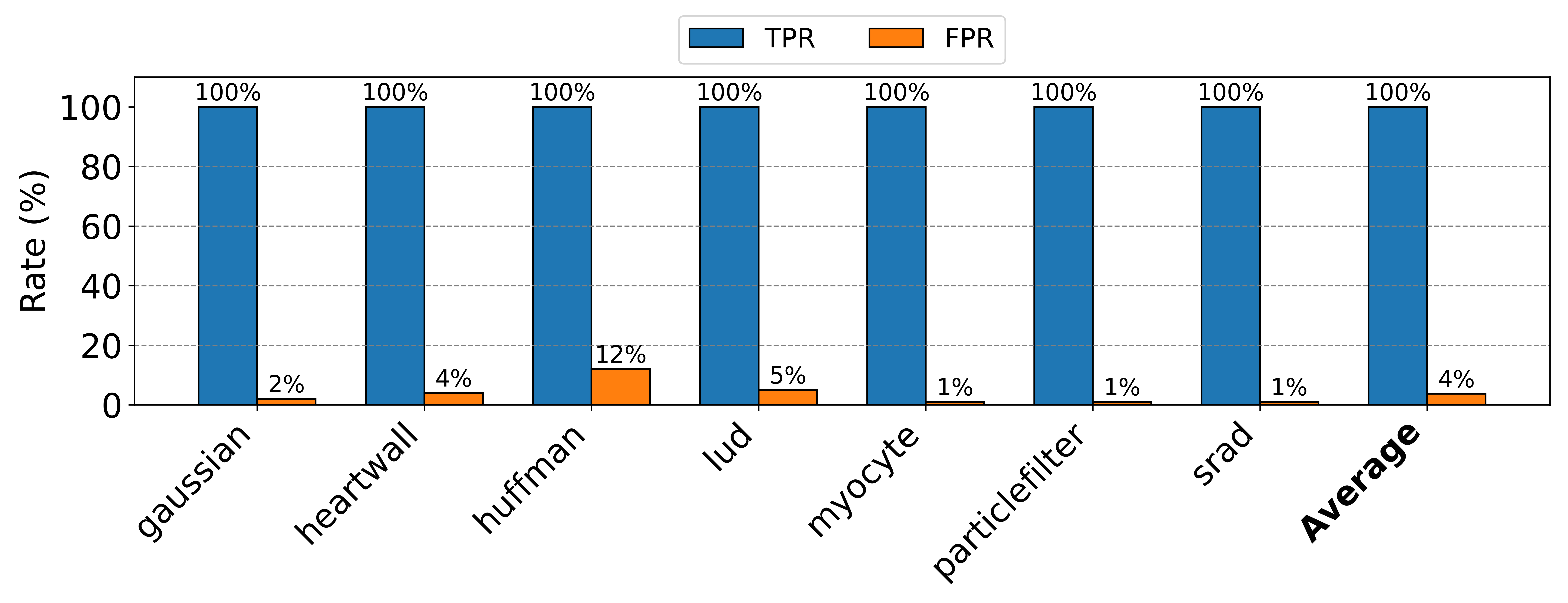}
    \caption{Performance of \sysname{} in monitoring DoS/Slow-down attacks.}
    \label{fig:attack_4_results}
    \vspace{-5mm}
\end{figure}

Figure~\ref{fig:attack_4_results} presents the results of \sysname{} on seven benchmarks from the Rodinia suite. \sysname{} achieves perfect detection (100\% TPR) across all benchmarks, highlighting its strong sensitivity to attack behavior. Most benchmarks show low FPRs, typically between 1\% and 5\%. However, \textit{huffman} exhibits a higher FPR of 12\%. Even so, the system delivers excellent overall performance, achieving an average TPR of 100\% and an average FPR of just 4\%.


\subsection{Robustness to Noise}

Since GPUs are designed for high parallelism and can run multiple programs concurrently, we evaluate how \sysname{} performs under kernel interference. To assess the impact of noise, we first collect 20 traces with only AlexNet running on the GPU. We then introduce two noise scenarios separately: concurrent execution of additional AlexNet models (self-noise), and concurrent execution of other benchmarks such as VecAdd (external noise). For each condition, we collect another 20 traces. We use a normalized DTW similarity score~\cite{salvador2007toward}, where values closer to 1 indicate higher similarity, to compare noisy traces against the baseline.

\begin{figure}[ht]
    \centering
    \includegraphics[width=0.8\linewidth]{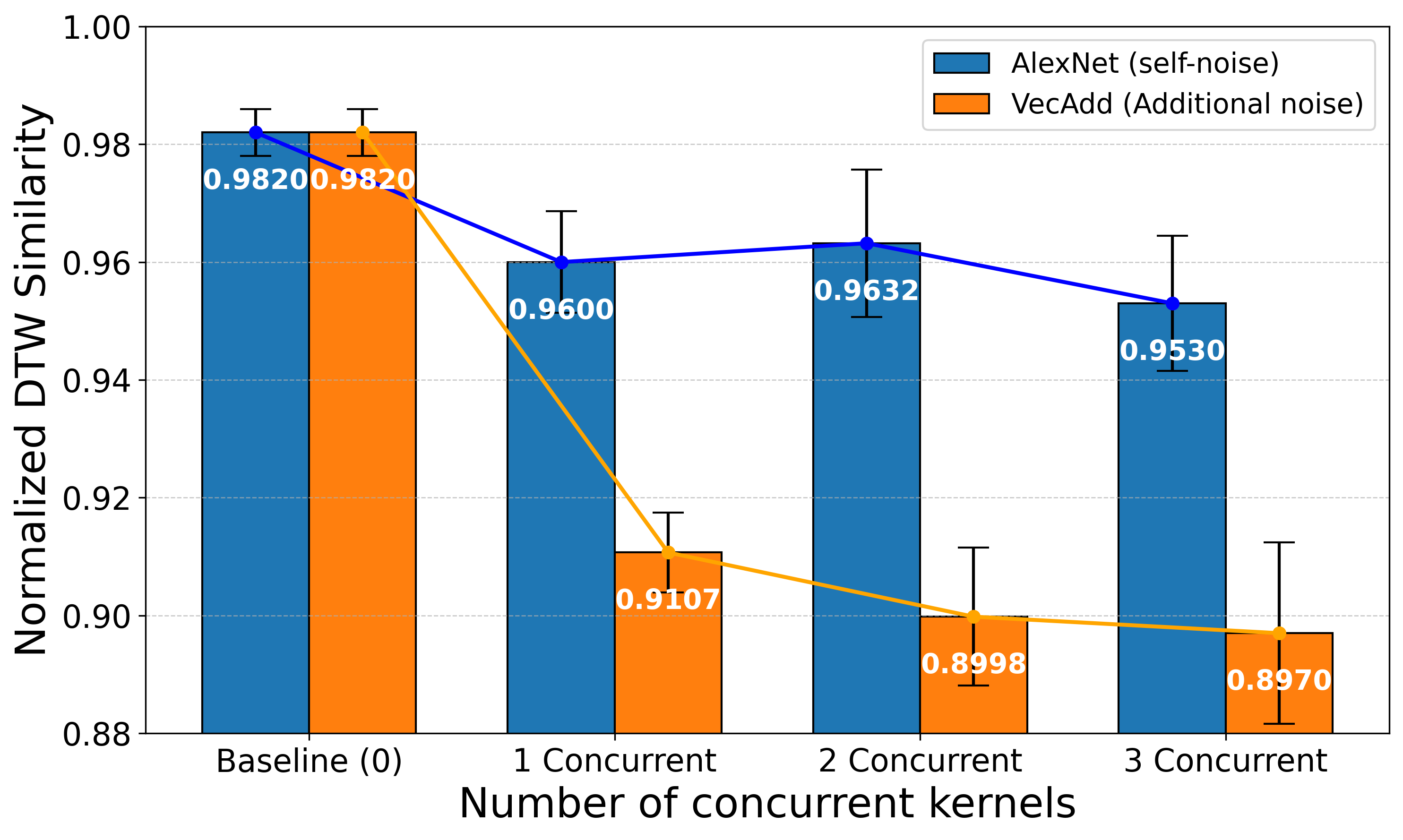}
    \caption{Impact of concurrent kernel noise (\sysname{}).}
    \label{fig:noise_alexnet}
\end{figure}

Figure~\ref{fig:noise_alexnet} shows the effect of concurrent kernel execution on normalized DTW similarity as measured by \sysname{}. We compare two scenarios: (1) AlexNet as self-noise, where the same model runs alongside itself, and (2) VecAdd as external noise, where a lightweight benchmark runs concurrently with AlexNet. In the baseline setting without concurrent kernels, the similarity score averages 0.9820. As the number of concurrent AlexNet kernels increases from one to three, similarity gradually declines. The drop is more pronounced in the VecAdd scenario, where the score falls to 0.8970. Although this value still indicates strong similarity, it may lead to a higher false positive rate. In comparison, the score under self-noise remains higher at 0.9530.

\subsection{Limitation of Existing PMU}
\label{subsec:limitation_PMU}

Even though \sysname{} performs well in most GPU benchmark scenarios, it depends heavily on NVIDIA’s existing GPU Performance Monitoring Unit (PMU). However, the current PMU design presents several limitations that may hinder accurate kernel validation and anomaly detection. In this section, we summarize the key limitations of the existing GPU PMU.

\noindent \textbf{Sampling rate.} The sampling rate of NVIDIA CUPTI is relatively low, which can lead to insufficient data for fast-executing kernels. As a result, the golden model may fail to capture enough information to validate kernel execution accurately.

\noindent \textbf{Events grouping.} CUPTI enforces fixed event groupings, preventing the simultaneous collection of certain event combinations. This limits the ability to fully observe kernel behavior during execution.

\noindent \textbf{Profiling overhead.} CUPTI-based profiling introduces measurable runtime overhead. This can distort the behavior of lightweight kernels and reduce the fidelity of the collected traces.

\noindent \textbf{Interference from concurrent kernels.} When multiple kernels run concurrently, event counters may overlap or interfere with one another, making it difficult to isolate and attribute events to specific kernels.

To address these limitations, we propose \hwname{}, a hardware-assisted framework for \textbf{\textit{in-GPU kernel validation}}, as described in Section~\ref{sec:gpu_centric_shadowscope}.

\section{GPU-Centric \sysname{}}
\label{sec:gpu_centric_shadowscope}

To address the challenges mentioned in Section~\ref{subsec:limitation_PMU}, we propose \hwname{}, a hardware-assisted framework for \textbf{\textit{in-GPU kernel validation}} that operates independently of the CPU. \hwname{} leverages the GPU’s existing hardware PMUs in conjunction with a dedicated on-chip validator to perform real-time validation of kernel execution. By localizing validation logic entirely within the GPU, \hwname{} minimizes performance overhead and improves parallelism. Unlike CUPTI-based solutions, it also avoids the need for privileged software components or driver-level modifications, thereby reducing the trusted computing base and easing deployment.

\subsection{\hwname{} Design}
\label{sec:hw_design}

\begin{figure}
    \centering
    \includegraphics[width=\linewidth]{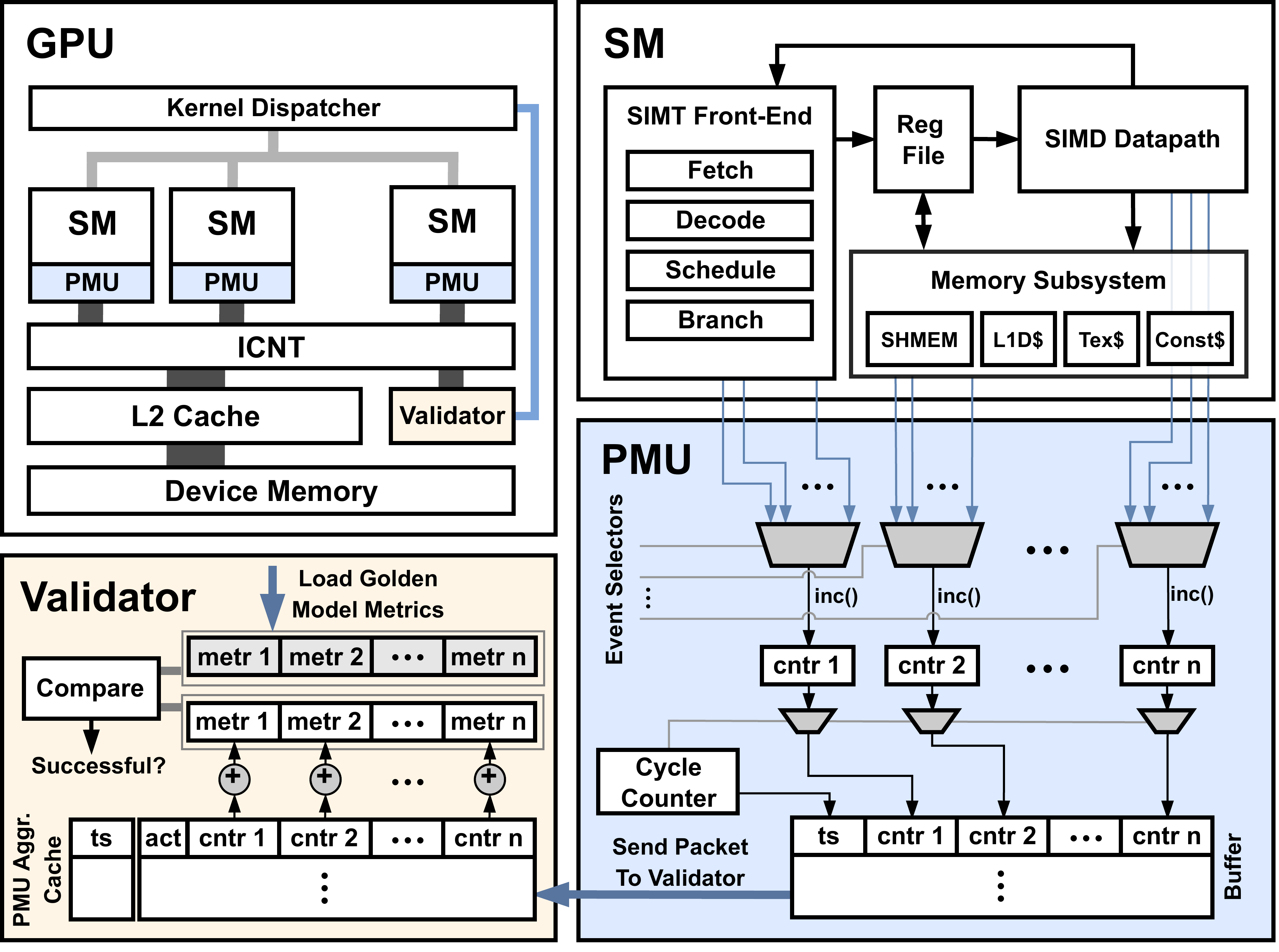}
    \caption{\hwname{} high-level architecture.}
    \label{fig:gpu_pmu_arch}
    \vspace{-5mm}
\end{figure}

We equip each SM with a local PMU capable of independently collecting performance data, as illustrated in Figure~\ref{fig:gpu_pmu_arch}. This is similar to \textit{Streaming Multiprocessor Performance Counter (SMPC)} model in NVIDIA GPUs~\cite{nvidia_nsight_perfctr, saiz2022top}. In addition, we introduce an on-chip Validator module integrated into the GPU’s Interconnection Network (\textit{ICNT}) alongside other SMs and memory partitions (including L2 cache slices). This Validator is responsible for analyzing and validating the performance samples collected by the local PMUs.

\subsubsection{Performance Monitoring Unit (PMU)}
\label{sec:pmu_design}

As illustrated in Figure~\ref{fig:gpu_pmu_arch}, each SM in the proposed design integrates a local PMU responsible for capturing microarchitectural events during kernel execution. The PMU receives a set of one-bit, cycle-wide signals from various SM components, each signaling the occurrence of specific events, such as instruction type counts, L1 data cache hits or misses, warp issues, or idle scheduler slots. These signals are routed to a set of configurable multiplexers, each of which selects one event signal to monitor. The selection logic is controlled via firmware-accessible registers, as commonly used by tools like NVIDIA Nsight~\cite{nvidia_nsight_perfctr} and CUPTI~\cite{cupti,cupti_doc}. Each multiplexer output is connected to a dedicated 32-bit up-counter that increments on every cycle the selected event signal is asserted. This straightforward yet effective design enables fine-grained, per-SM event tracking with low overhead and minimal disruption to kernel execution.

At the end of each sampling window, determined either by a firmware-configurable cycle counter register or by the end of the kernel execution, the PMU freezes all active counters and stores their values in a buffer entry, along with the corresponding timestamp or cycle count (\textit{ts}), and resets the counters to zero in preparation for the next sampling window. In standard profiling mode (i.e., when kernel validation is disabled), a lightweight dedicated DMA engine transfers these buffer entries into a ring buffer located in the GPU's global device memory. This data can subsequently be used by existing profiling tools for performance analysis and reporting, or leveraged by our framework to construct a golden model representing the performance profile of a benign kernel execution. When kernel validation is enabled, however, \hwname{} redirects the buffer entries to the Validator module over the GPU interconnection network, utilizing idle or underutilized bandwidth to minimize interference with the primary workload.

\subsubsection{Validator Module}
\label{sec:validator_design}

As performance samples are received from active PMUs, the Validator first consults a small PMU aggregation cache, which tracks the accumulation of performance counter values for each sampling window (identified by a timestamp) across active PMUs. Upon receiving a packet, the Validator uses the timestamp as a cache tag to look up the corresponding entry. If no entry exists, a new one is created, and the active PMU count (\textit{act}) field is initialized based on the number of active PMUs for the currently executing kernel, as specified by the kernel dispatcher. If an entry is found, the incoming metric values are added to the existing values in the cache entry and the act count is decremented to reflect the arrival of data from one of the active PMUs. Once the \textit{act} count reaches zero, indicating that all expected PMU samples for that window have been received, the Validator reads the accumulated values and updates a set of internal registers representing the total monitored metrics for that sampling window. These aggregated metrics are then compared against the preloaded golden model values by computing the distance between corresponding metric pairs, as illustrated in Figure~\ref{fig:gpu_pmu_arch}. If the distance for these metrics exceeds a pre-defined threshold, tuned according to the sensitivity and semantics of the selected metrics, the kernel is flagged as potentially malicious. In such cases, the Validator signals the kernel dispatcher to stop the kernel execution and notifies the CPU through the GPU driver to initiate appropriate mitigation steps. If the comparison passes, indicating no malicious behavior, the Validator proceeds to monitor subsequent sampling windows.

\subsection{\hwname{} Mechanism}
\label{sec:hw_mechanism}

\begin{figure}[tbh]
    \centering
    \includegraphics[width=\linewidth]{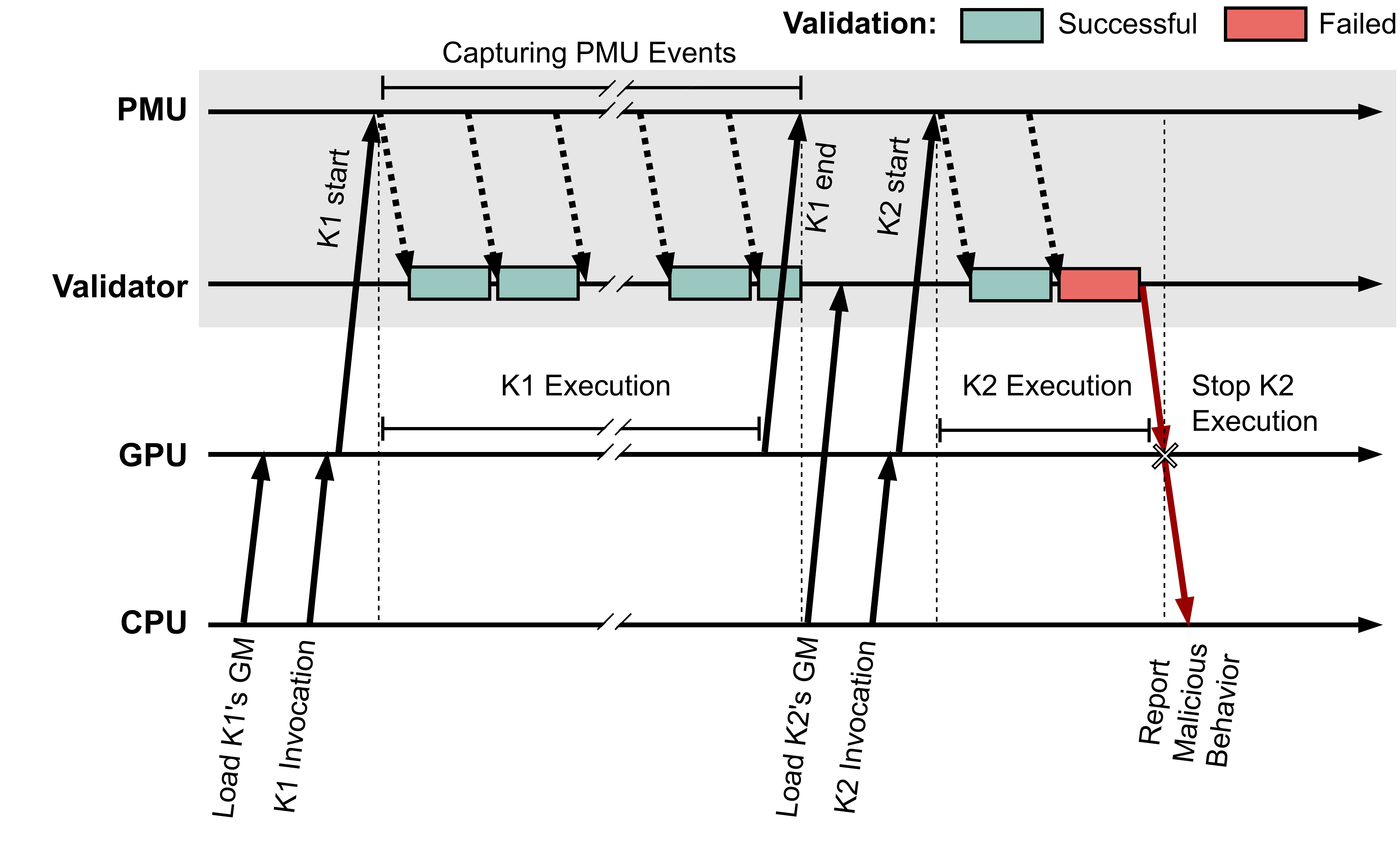}
    \caption{An overview of the \hwname{} mechanism for in-GPU kernel validation.}
    \label{fig:gpu_pmu_mechanism}
    \vspace{-5mm}
\end{figure}

Figure~\ref{fig:gpu_pmu_mechanism} illustrates the operational flow of the proposed monitoring and validation mechanism in \hwname{}. At the time of kernel invocation, the golden model corresponding to Kernel 1 (K1) is loaded into the GPU’s global device memory, where it is later accessed by the Validator for runtime comparison. Kernel execution begins when the kernel dispatcher assigns kernel to a set of active SMs. Upon dispatch, the dispatcher also notifies the Validator of the kernel launch and the associated active SMs. Each of these SMs activates its local PMU, which begins collecting microarchitectural performance events specific to the executing kernel.

During execution, each local PMU accumulates event samples and transmits them to the Validator at the end of every predefined sampling window, or upon kernel completion. The Validator then performs runtime validation by aggregating the metrics from all active PMUs and comparing them against the expected values from the golden model for the corresponding sampling window (as illustrated by the colored boxes on the Validator timeline in Figure~\ref{fig:gpu_pmu_mechanism}). If the collected samples align with the golden model within acceptable thresholds, the kernel is classified as benign, and execution continues without interruption (as indicated by the green boxes in the timeline).

In contrast, if discrepancies are detected, such as deviations in control flow or memory access behavior, as in the case of Kernel 2 (K2), the Validator flags the violation by identifying significant deviations  (as indicated by the red box in the timeline). It then signals the GPU to halt execution of the kernel and reports the anomaly and the offending kernel (K2) to the CPU via the GPU driver. 

\subsection{Simulation Methodology}
\label{sec:sim_methodology}

To implement and evaluate our proposed design, we extend GPGPU-Sim v4.0~\cite{gppgusim}, a cycle-accurate simulator for NVIDIA GPU architectures. Our proof-of-concept is modeled on a small GPU configuration based on the NVIDIA Fermi architecture and the GTX480 platform. Detailed simulation parameters are provided in Table~\ref{tab:sim_config}. For evaluation, we use four CUDA Samples benchmarks, \textit{vectorAdd}, \textit{matrixMul}, \textit{histogram}, and \textit{bitonicSort}, along with two DNN models, \textit{AlexNet} and \textit{CifarNet}. These workloads are used in Sections~\ref{sec:hw_security_eval} and~\ref{sec:hw_performance_eval} to assess the effectiveness of the validation and detection mechanisms, as well as to measure performance overhead.

\begin{table}[h]
    \centering
    \caption{Simulator configuration parameters.}
    \label{tab:sim_config}
    \renewcommand{\arraystretch}{1.2}
    \resizebox{0.9\linewidth}{!}{
    \begin{tabular}{|l|p{5cm}|}
    \hline
    \textbf{Parameter} & \textbf{Specification} \\
    \hline \hline
    \textbf{Number of SMs} & 15 \\
    \hline
    \textbf{SM Configuration} & Warps/SM = 48, Schedulers/SM = 2, Warps/Scheduler = 16, Register File/SM = 32 KB \\
    \hline
    \textbf{Execution Units} & 2 SPs, 1 SFUs \\
    \hline
    \textbf{Shared Memory} & 48 kB, latency = 26 cycles\\
    \hline
    \textbf{L1 Data Cache} & 16 KB, 8-sets, 16-ways\\
    \hline
    \textbf{L2 Cache} & 786 KB, 64-sets, 16-ways, 6-banks\\
    \hline
    \textbf{Min Access Latency} & L1 = 35 cycles, L2 = 120 cycles \\
    \hline
    \textbf{Memory Model} & GDDR5, latency = 220 cycles \\
    \hline
    \textbf{Frequency} & SM,ICNT,L2:700 MHz, MEM:924 MHz \\
    \hline
    \end{tabular}}

\end{table}



\subsection{Evaluation Results}
\subsubsection{Security Evaluation}
\label{sec:hw_security_eval}

We evaluate the detection effectiveness of \hwname{} in the presence of two known GPU attacks: buffer overflow and mind-control attacks.
To quantify deviations in kernel execution, we employ normalized \textit{Dynamic Time Warping} (DTW) scores to measure the similarity between benign and attack execution traces of the same kernel, captured using custom hardware PMUs and analyzed by the Validator proposed in \hwname{}. 

\noindent \textbf{Kernel Deviation Attacks.} 
Table~\ref{tb:gpgpu_attack_1} presents the average and standard deviation of normalized DTW similarity scores between benign and attack kernel executions across four benchmarks. Lower scores indicate greater divergence from expected behavior; typically, a score below 0.1 suggests low similarity. Among the benchmarks, \textit{matMul} shows the highest similarity score (0.0876), indicating minimal deviation, while \textit{bitonicSort} and \textit{vecAdd} exhibit the lowest scores (0.0178 and 0.0181, respectively), reflecting more substantial deviations caused by the attacks. Notably, \textit{vecAdd} yields a standard deviation of 0, as it contains only a single executed kernel. These results demonstrate the effectiveness of our approach in identifying anomalous executions.

\begin{table}[ht]
\centering
\small
\caption{Normalized DTW similarity scores between benign and attack traces.}
\renewcommand{\arraystretch}{1.2}
\label{tb:gpgpu_attack_1}
\begin{tabular}{|l|cc|}

\hline
\multirow{2}{*}{\textbf{Benchmarks}} & \multicolumn{2}{c|}{\textbf{Similarity score}}                      \\ \cline{2-3} 
                                     & \multicolumn{1}{c|}{\textbf{average}} & \textbf{std}                \\ \hline \hline
bitonicSort                          & \multicolumn{1}{c|}{0.0178}           & 0.0151                      \\ \hline
histogram                            & \multicolumn{1}{c|}{0.0427}           & 0.0315                      \\ \hline
matMul                               & \multicolumn{1}{c|}{0.0876}           & \multicolumn{1}{l|}{0.0088} \\ \hline
vecAdd                               & \multicolumn{1}{c|}{0.0181}           & \multicolumn{1}{l|}{-}      \\ \hline
\end{tabular}
\end{table}

\noindent \textbf{Mind Control Attacks.} In this attack, we evaluate \textit{AlexNet} and \textit{CifarNet}. For AlexNet, using \hwname{}, we collect 10 segments of traces during benign execution, with each segment corresponding to the execution of one layer or kernel. In the attack scenario, one layer is skipped, resulting in only 9 segments being collected. This allows us to detect that the second layer was skipped. Similarly, for CifarNet, we also detect that the second layer is missing, indicating a successful identification of the mind control attack.


\subsubsection{Performance Evaluation}
\label{sec:hw_performance_eval}

\begin{figure}[ht]
    \centering
    \includegraphics[width=\linewidth]{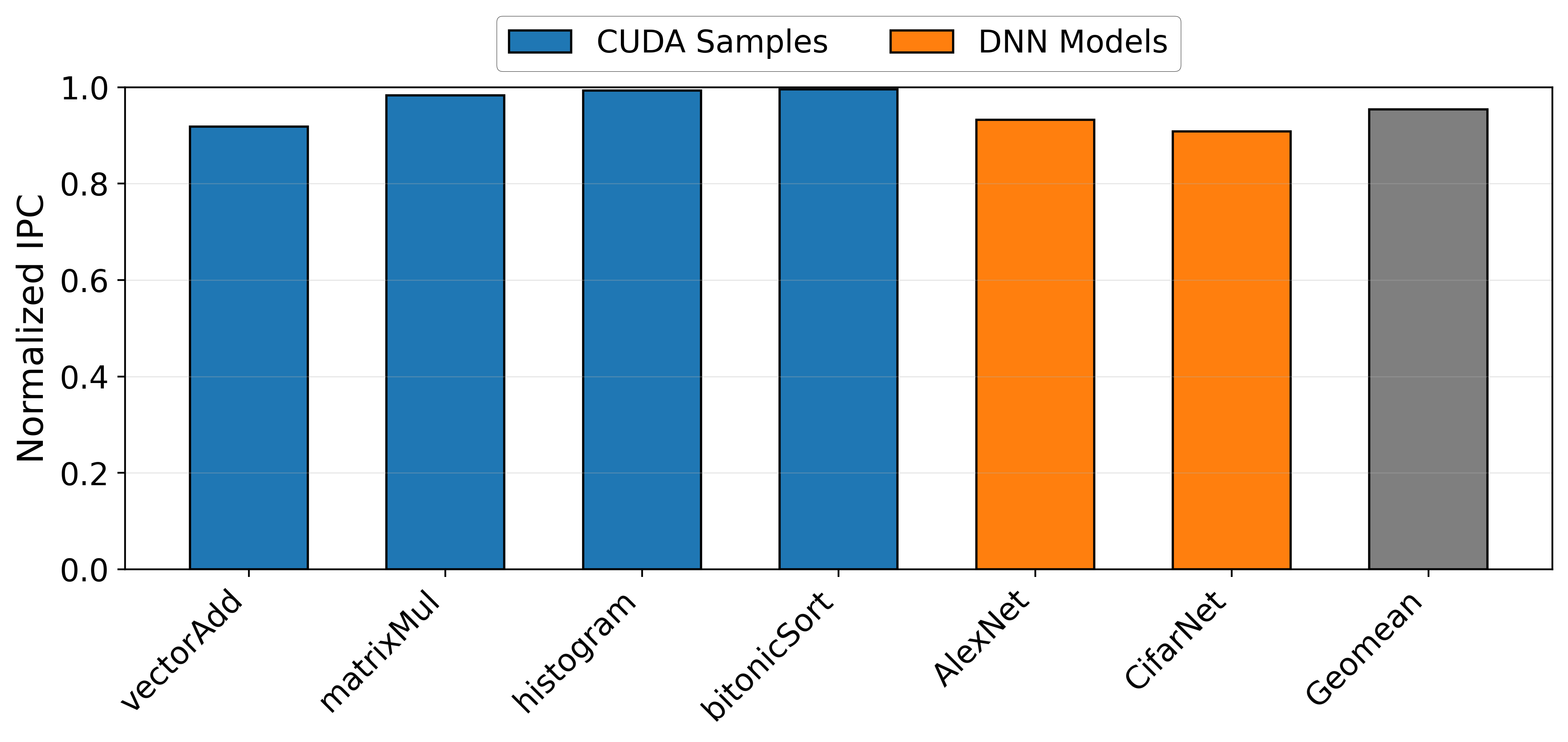}
    \caption{Performance normalized to baseline GPU.}
    \label{fig:hw_perf}
\end{figure}

Figure~\ref{fig:hw_perf} presents the performance overhead introduced by \hwname{} compared to the baseline GPU architecture across the evaluated benchmarks. Since both PMU event monitoring and kernel validation occur off the critical path of kernel execution, their impact on performance is minimal. The primary source of overhead stems from the transfer of collected samples via the GPU’s shared interconnection network by the local SM PMUs, and the loading of golden model samples by the Validator through shared memory channels into its local fetch buffer. This leads to contention on the interconnect bandwidth and memory channels, particularly in applications that heavily utilize these shared resources, such as memory-intensive workloads like DNN models. The results show that \hwname{} incurs a geometric mean performance overhead of 4.6\% in terms of normalized IPC compared to the baseline, with the lowest overhead observed in bitonicSort at 0.4\%, and the highest in CifarNet at 9.2\%. 

\subsubsection{Hardware Complexity Evaluation}
\label{sec:hw_complexity_eval}

For our hardware cost evaluation, each PMU is modeled with eight 32-bit counter registers and eight 8-to-1 multiplexers to monitor different microarchitectural events per counter. A 32-bit cycle counter generates timestamps for sampling windows. Each PMU includes an output buffer with eight 36-byte entries (288 bytes total) to store timestamps and associated metric values. The Validator’s fetch buffer mirrors this structure but at half the size, four entries totaling 144 bytes, to hold golden model metrics loaded from the global device memory of the GPU. The PMU aggregation buffer is implemented as a simple cache with 33-byte blocks containing eight metrics, an 8-bit active-SM counter value, and a 32-bit timestamp as the tag. Aggregation is performed using eight 32-bit adders, while the Validator’s comparison logic consists of eight subtractors, eight magnitude comparators, and a single threshold register to compute deviations from the golden model.

We evaluated SRAM-based structures using CACTI~\cite{CACTI} using 40nm technology, the same technology used by AccelWattch~\cite{kandiah2021accelwattch}, for modeling the NVIDIA GTX480 GPU. Key logic components, including counters, multiplexers, and comparators, were implemented in Verilog and synthesized using Synopsys Design Compiler 2017.09. Table~\ref{tbl:hardware_cost} reports the overhead of a single PMU relative to an SM, the cumulative overhead of all PMUs relative to the full GPU, and the Validator’s cost as a global module relative to the full GPU. 

\begin{table}[t]
\centering
\scriptsize
\caption{\small Area and power overhead (percentage increase from baseline) of integrating PMUs into each SM (reported per SM and per GPU) and the Validator (reported relative to the entire GPU).}
\label{tbl:hardware_cost}
\resizebox{0.9\linewidth}{!}{
\begin{tabular}{@{\hspace{2mm}}l@{\hspace{2mm}}l@{\hspace{2mm}}c@{\hspace{2mm}}c@{\hspace{2mm}}c}
\textbf{Component} & \textbf{Metric}
& \textbf{Per SM} 
& \textbf{Per GPU} \\
\midrule
\multirow{3}{*}{PMU}
    & Area Overhead (\%)  & 0.03  & 0.02  \\
    & Static Power (\%)  & 0.07  & 0.13  \\
    & Dynamic Power (\%)  & 0.28  & 0.26  \\
\midrule
\multirow{3}{*}{Validator}
    & Area Overhead (\%)  & -  & 0.004  \\
    & Static Power (\%)  & -  & 0.01  \\
    & Dynamic Power (\%)  & -  & 0.02  \\
\midrule
\multirow{3}{*}{}  
    & Total Area Overhead (\%) & & 0.03   \\
    & Total Static Power (\%) & & 0.13  \\
    & Total Dynamic Power (\%) & & 0.28   \\
\bottomrule
\end{tabular}
}
\vspace{-0.2in}
\end{table}

\section{Related Work}
\label{sec:related_work}

\noindent\textbf{Golden model validation.} A variety of approaches have been proposed for hardware trojan detection using golden models, which are constructed from trusted hardware side-channel data collected through physical sources such as power, leakage current, and temperature~\cite{current_gm, wirelss_gm, power_gm, temp_pow_gm}.

\noindent\textbf{Attacks detection and software validation using hardware performance counter (HPC).}
HPCs collected from the PMU have been widely used to detect malware and side-channel attacks by identifying deviations in software behavior.
Demme at al.~\cite{hpc_columbia} explored the feasibility of using HPC for malware and side channel detection in both ARM and Intel CPUs. 
Lee et al.~\cite{spectre_hpc} proposed using HPC to detect Spectre attacks by randomly selecting detectors, feature sets, and sampling periods to improve robustness against sophisticated threats.
Khasawneh et al.~\cite{rhmd_hpc} focused on detecting evasive malware by designing a system with multiple detectors and randomly selecting among them, making it difficult for attackers to evade detection.
Most prior work using HPCs has focused on CPU-based systems and malware detection, with limited emphasis on software validation and little attention to attacks targeting GPU kernels.

\noindent\textbf{Control flow attacks targeting GPU kernels.}
Park et al.~\cite{mind_control_attack} proposed reducing the prediction accuracy of DNN models by accessing arbitrary memory locations used by targeted ML models. The attack is based on a buffer overflow attack which is used to inject random code to lower the accuracy of ML models. 
Recently, Guo et al.~\cite{guo2024gpu} investigated buffer overflow vulnerabilities in NVIDIA GPUs and demonstrated traditional code injection attacks and ROP-style attacks are possible on GPUs. 
Roels et al.~\cite{GPU_code_reuse_attack_2025} adapted CPU-style code reuse attacks to NVIDIA GPUs, enabling the discovery of new ROP gadgets and the construction of Turing-complete ROP attacks on GPUs.


\noindent\textbf{Exploiting PMU as a side channel in GPUs.}
By collecting data from GPU PMU during victim kernel execution, attackers can leak secret information about the victim. 
Naghibijouybari et al.~\cite{naghibijouybari2018rendered} exploited performance counters reflecting shared resource usage between victim and attacker to perform a series of side-channel attacks.
Wei et al.~\cite{wei2020leaky} exploited data collected from GPU PMU to monitor context switches during DNN model training to learn targeted model layers and parameters. 
Wang et al.~\cite{wang2022_pmu} also targeted DNN to perform model extraction attacks and mainly targeted events related to a unified memory management system which they found to have higher effectiveness.

\noindent\textbf{Software and hardware defenses in GPUs.} Hardware and software-based approaches propose the implementation of TEEs in GPUs. 
Volos et al.~\cite{graviton} to secure GPU kernels with no modification to CPU. Their approach only changes the command processor within GPU for TEE support. 
Jang et al.~\cite{jang_hix} also proposed TEE support for GPU. Their design requires modifications to the  I/O interconnect and changes to the GPU driver to be included within CPU TEE. 
NVIDIA H100 includes hardware support for confidential computing (CC) targeting virtualized environment~\cite{h100_cc}. To isolate a virtual machine (VM), H100 CC requires support for TEE at the CPU side as well. TEE at the CPU side can be achieved through Intel TDX~\cite{intel_tdx}, AMD SEV-SNP~\cite{amd_sev}, or ARM CCA~\cite{arm_cca}. 


\noindent\textbf{Software attestation using side channels.}
Existing work also proposed using side channels such as power and electromagnetic signals for good use such as software attestation. Side channel data of program execution based on electromagnetic signals are used to detect deviations in program execution~\cite{sehatbakhsh2019emma,nazari2017eddie}. This research targeted embedded systems and assumes predictable execution; ShadowScope extends the range of such approaches using the idea of composable verification.  A line of research uses performance counters on CPUs to classify programs as benign or malicious~\cite{ozsoy-16,rhmd_hpc,samira-22}; however, these approaches provide a weaker protection since they do not verify the correct execution of a program.

\section{Concluding Remarks}
\label{sec:conclusion}
We present \sysname{}, a robust monitoring and validation framework for GPU kernel execution based on composable golden models derived from side-channel signals. By capturing modular and repeatable execution patterns, \sysname{} effectively detects both anomalous behavior within trusted kernels and signs of kernel compromise. It detects four types of attacks with up to 100\% true positive rate and low false positive rates. To further reduce noise sensitivity and performance overhead, we introduce \hwname{}, a lightweight hardware-assisted mechanism for on-chip validation, achieving accurate detection with only 4.6\% performance overhead.



\bibliographystyle{plainurl}
\bibliography{refs}

\begin{thebibliography}{10}

\bibitem{gpu_book}
Tor~M. Aamodt, Wilson Wai~Lun Fung, and Timothy~G. Rogers.
\newblock General-purpose graphics processor architectures, 2018.

\bibitem{samira-22}
Samira~Mirbagher Ajorpaz, Daniel Moghimi, Jeffrey~Neal Collins, Gilles Pokam, Nael Abu-Ghazaleh, and Dean Tullsen.
\newblock Evax: Towards a practical, pro-active \& adaptive architecture for high performance \& security.
\newblock In {\em 2022 55th IEEE/ACM International Symposium on Microarchitecture (MICRO)}, pages 1218--1236. IEEE, 2022.

\bibitem{gpu_intro}
Amazon.
\newblock {What is a GPU?}
\newblock https://aws.amazon.com/what-is/gpu/.
\newblock Last accessed on: 07/31/2024.

\bibitem{amd_sev}
{AMD}.
\newblock {AMD Secure Encrypted Virtualization (SEV)}.
\newblock https://www.amd.com/en/developer/sev.html.
\newblock Last accessed on: 05/25/2025.

\bibitem{amd_GPUPerfAPI}
AMD.
\newblock {GPUPerfAPI}.
\newblock https://gpuopen.com/gpuperfapi/.
\newblock Last accessed on: 07/27/2024.

\bibitem{arm_cca}
{arm}.
\newblock {Arm Confidential Compute Architecture}.
\newblock https://www.arm.com/architecture/security-features/arm-confidential-compute-architecture.
\newblock Last accessed on: 05/25/2025.

\bibitem{gpu_simt}
{Caroline Collange}.
\newblock {GPU architecture: Revisiting the SIMT execution model}.
\newblock http://www.irisa.fr/alf/downloads/collange/cours/hpca2020\_gpu\_0.pdf.
\newblock Last accessed on: 07/31/2024.

\bibitem{cathislapd}
Alexander Cathis, Mulong Luo, Mohit Tiwari, and Andreas Gerstlauer.
\newblock Lapd: Lifecycle-aware power-based malware detection.
\newblock {\em IEEE International Symposium on Hardware Oriented Security and Trust (HOST)}, 2025.

\bibitem{rodinia1}
Shuai Che, Michael Boyer, Jiayuan Meng, David Tarjan, Jeremy~W. Sheaffer, Sang-Ha Lee, and Kevin Skadron.
\newblock Rodinia: A benchmark suite for heterogeneous computing.
\newblock In {\em 2009 IEEE International Symposium on Workload Characterization (IISWC)}, pages 44--54, 2009.
\newblock \href {https://doi.org/10.1109/IISWC.2009.5306797} {\path{doi:10.1109/IISWC.2009.5306797}}.

\bibitem{hpc_columbia}
John Demme, Matthew Maycock, Jared Schmitz, Adrian Tang, Adam Waksman, Simha Sethumadhavan, and Salvatore Stolfo.
\newblock On the feasibility of online malware detection with performance counters.
\newblock In {\em Proceedings of the 40th Annual International Symposium on Computer Architecture}, ISCA '13, page 559–570, New York, NY, USA, 2013. Association for Computing Machinery.
\newblock \href {https://doi.org/10.1145/2485922.2485970} {\path{doi:10.1145/2485922.2485970}}.

\bibitem{gpu_buf_of}
Bang Di, Jianhua Sun, and Hao Chen.
\newblock A study of overflow vulnerabilities on gpus.
\newblock In {\em Network and Parallel Computing: 13th IFIP WG 10.3 International Conference, NPC 2016, Xi'an, China, October 28-29, 2016, Proceedings}, page 103–115, Berlin, Heidelberg, 2016. Springer-Verlag.
\newblock \href {https://doi.org/10.1007/978-3-319-47099-3_9} {\path{doi:10.1007/978-3-319-47099-3_9}}.

\bibitem{eastman2017openmm}
Peter Eastman, Jason Swails, John~D Chodera, Robert~T McGibbon, Yutong Zhao, Kyle~A Beauchamp, Lee-Ping Wang, Andrew~C Simmonett, Matthew~P Harrigan, Chaya~D Stern, et~al.
\newblock Openmm 7: Rapid development of high performance algorithms for molecular dynamics.
\newblock {\em PLoS computational biology}, 13(7):e1005659, 2017.

\bibitem{buf_of_cgo}
Christopher Erb, Mike Collins, and Joseph~L. Greathouse.
\newblock {Dynamic buffer overflow detection for GPGPUs}.
\newblock In {\em 2017 IEEE/ACM International Symposium on Code Generation and Optimization (CGO)}, pages 61--73, 2017.
\newblock \href {https://doi.org/10.1109/CGO.2017.7863729} {\path{doi:10.1109/CGO.2017.7863729}}.

\bibitem{frigo2018grand}
Pietro Frigo, Cristiano Giuffrida, Herbert Bos, and Kaveh Razavi.
\newblock Grand pwning unit: Accelerating microarchitectural attacks with the gpu.
\newblock In {\em 2018 ieee symposium on security and privacy (sp)}, pages 195--210. IEEE, 2018.

\bibitem{guo2024gpu}
Yanan Guo, Zhenkai Zhang, and Jun Yang.
\newblock {GPU Memory Exploitation for Fun and Profit}.
\newblock In {\em 33rd USENIX Security Symposium (USENIX Security 24)}, pages 4033--4050, 2024.

\bibitem{he2015model}
Sunny~L He, Natalie~H Roe, Evan Wood, Noel~M Nachtigal, and Jovana Helms.
\newblock Model of the product development lifecycle.
\newblock Technical report, Sandia National Lab.(SNL-NM), Albuquerque, NM (United States), 2015.

\bibitem{temp_pow_gm}
Kangqiao Hu, Abdullah~Nazma Nowroz, Sherief Reda, and Farinaz Koushanfar.
\newblock High-sensitivity hardware trojan detection using multimodal characterization.
\newblock In {\em 2013 Design, Automation \& Test in Europe Conference \& Exhibition (DATE)}, pages 1271--1276, 2013.
\newblock \href {https://doi.org/10.7873/DATE.2013.263} {\path{doi:10.7873/DATE.2013.263}}.

\bibitem{hu2020deepsniffer}
Xing Hu, Ling Liang, Shuangchen Li, Lei Deng, Pengfei Zuo, Yu~Ji, Xinfeng Xie, Yufei Ding, Chang Liu, Timothy Sherwood, et~al.
\newblock Deepsniffer: A dnn model extraction framework based on learning architectural hints.
\newblock In {\em Proceedings of the Twenty-Fifth International Conference on Architectural Support for Programming Languages and Operating Systems}, pages 385--399, 2020.

\bibitem{intel_tdx}
{intel}.
\newblock {Intel® Trust Domain Extensions}.
\newblock https://www.intel.com/content/www/us/en/developer/tools/trust-domain-extensions/overview.html.
\newblock Last accessed on: 05/25/2025.

\bibitem{ivanov2023sage}
Andrei Ivanov, Benjamin Rothenberger, Arnaud Dethise, Marco Canini, Torsten Hoefler, and Adrian Perrig.
\newblock $\{$SAGE$\}$: Software-based attestation for $\{$GPU$\}$ execution.
\newblock In {\em 2023 USENIX Annual Technical Conference (USENIX ATC 23)}, pages 485--499, 2023.

\bibitem{jang_hix}
Insu Jang, Adrian Tang, Taehoon Kim, Simha Sethumadhavan, and Jaehyuk Huh.
\newblock Heterogeneous isolated execution for commodity gpus.
\newblock In {\em Proceedings of the Twenty-Fourth International Conference on Architectural Support for Programming Languages and Operating Systems}, ASPLOS '19, page 455–468, New York, NY, USA, 2019. Association for Computing Machinery.
\newblock \href {https://doi.org/10.1145/3297858.3304021} {\path{doi:10.1145/3297858.3304021}}.

\bibitem{kande2022thehuzz}
Rahul Kande, Addison Crump, Garrett Persyn, Patrick Jauernig, Ahmad-Reza Sadeghi, Aakash Tyagi, and Jeyavijayan Rajendran.
\newblock $\{$TheHuzz$\}$: Instruction fuzzing of processors using $\{$Golden-Reference$\}$ models for finding $\{$Software-Exploitable$\}$ vulnerabilities.
\newblock In {\em 31st USENIX Security Symposium (USENIX Security 22)}, pages 3219--3236, 2022.

\bibitem{kandiah2021accelwattch}
Vijay Kandiah, Scott Peverelle, Mahmoud Khairy, Junrui Pan, Amogh Manjunath, Timothy~G Rogers, Tor~M Aamodt, and Nikos Hardavellas.
\newblock Accelwattch: A power modeling framework for modern gpus.
\newblock In {\em MICRO-54: 54th Annual IEEE/ACM International symposium on microarchitecture}, pages 738--753, 2021.

\bibitem{gppgusim}
Mahmoud Khairy, Zhesheng Shen, Tor~M Aamodt, and Timothy~G Rogers.
\newblock Accel-sim: An extensible simulation framework for validated gpu modeling.
\newblock In {\em 2020 ACM/IEEE 47th Annual International Symposium on Computer Architecture (ISCA)}, pages 473--486. IEEE, 2020.

\bibitem{rhmd_hpc}
Khaled~N. Khasawneh, Nael Abu-Ghazaleh, Dmitry Ponomarev, and Lei Yu.
\newblock Rhmd: Evasion-resilient hardware malware detectors.
\newblock In {\em 2017 50th Annual IEEE/ACM International Symposium on Microarchitecture (MICRO)}, pages 315--327, 2017.

\bibitem{current_gm}
Charles Lamech, James Aarestad, Jim Plusquellic, Reza Rad, and Kanak Agarwal.
\newblock Rebel and tdc: Two embedded test structures for on-chip measurements of within-die path delay variations.
\newblock In {\em 2011 IEEE/ACM International Conference on Computer-Aided Design (ICCAD)}, pages 170--177, 2011.
\newblock \href {https://doi.org/10.1109/ICCAD.2011.6105322} {\path{doi:10.1109/ICCAD.2011.6105322}}.

\bibitem{lee2022securing}
Jaewon Lee, Yonghae Kim, Jiashen Cao, Euna Kim, Jaekyu Lee, and Hyesoon Kim.
\newblock Securing gpu via region-based bounds checking.
\newblock In {\em Proceedings of the 49th Annual International Symposium on Computer Architecture}, pages 27--41, 2022.

\bibitem{levi2020ask}
Itamar Levi, Davide Bellizia, David Bol, and Fran{\c{c}}ois-Xavier Standaert.
\newblock Ask less, get more: Side-channel signal hiding, revisited.
\newblock {\em IEEE Transactions on Circuits and Systems I: Regular Papers}, 67(12):4904--4917, 2020.

\bibitem{spectre_hpc}
Congmiao Li and Jean-Luc Gaudiot.
\newblock Detecting spectre attacks using hardware performance counters.
\newblock {\em IEEE Transactions on Computers}, 71(6):1320--1331, 2022.
\newblock \href {https://doi.org/10.1109/TC.2021.3082471} {\path{doi:10.1109/TC.2021.3082471}}.

\bibitem{CACTI}
Sheng Li, Ke~Chen, Jung~Ho Ahn, Jay~B Brockman, and Norman~P Jouppi.
\newblock Cacti-p: Architecture-level modeling for sram-based structures with advanced leakage reduction techniques.
\newblock In {\em 2011 IEEE/ACM International Conference on Computer-Aided Design (ICCAD)}, pages 694--701. IEEE, 2011.

\bibitem{lin2025gpuhammer}
Chris~S Lin, Joyce Qu, and Gururaj Saileshwar.
\newblock Gpuhammer: Rowhammer attacks on gpu memories are practical.
\newblock {\em arXiv preprint arXiv:2507.08166}, 2025.

\bibitem{liu2016code}
Yannan Liu, Lingxiao Wei, Zhe Zhou, Kehuan Zhang, Wenyuan Xu, and Qiang Xu.
\newblock On code execution tracking via power side-channel.
\newblock In {\em Proceedings of the 2016 ACM SIGSAC conference on computer and communications security}, pages 1019--1031, 2016.

\bibitem{wirelss_gm}
Yu~Liu, Yier Jin, and Yiorgos Makris.
\newblock Hardware trojans in wireless cryptographic ics: Silicon demonstration \& detection method evaluation.
\newblock In {\em 2013 IEEE/ACM International Conference on Computer-Aided Design (ICCAD)}, pages 399--404, 2013.
\newblock \href {https://doi.org/10.1109/ICCAD.2013.6691149} {\path{doi:10.1109/ICCAD.2013.6691149}}.

\bibitem{mangard2004hardware}
Stefan Mangard.
\newblock Hardware countermeasures against dpa--a statistical analysis of their effectiveness.
\newblock In {\em Topics in Cryptology--CT-RSA 2004: The Cryptographers’ Track at the RSA Conference 2004, San Francisco, CA, USA, February 23-27, 2004, Proceedings}, pages 222--235. Springer, 2004.

\bibitem{llama}
{Meta}.
\newblock {Llama}.
\newblock https://llama.meta.com.
\newblock Last accessed on: 07/31/2024.

\bibitem{miele2016buffer}
Andrea Miele.
\newblock Buffer overflow vulnerabilities in cuda: a preliminary analysis.
\newblock {\em Journal of Computer Virology and Hacking Techniques}, 12:113--120, 2016.

\bibitem{molina2007functional}
A~Molina and Oswaldo Cadenas.
\newblock Functional verification: Approaches and challenges.
\newblock {\em Latin American applied research}, 37(1):65--69, 2007.

\bibitem{mutlu2019rowhammer}
Onur Mutlu and Jeremie~S Kim.
\newblock Rowhammer: A retrospective.
\newblock {\em IEEE Transactions on Computer-Aided Design of Integrated Circuits and Systems}, 39(8):1555--1571, 2019.

\bibitem{naghibijouybari2017constructing}
Hoda Naghibijouybari, Khaled~N Khasawneh, and Nael Abu-Ghazaleh.
\newblock Constructing and characterizing covert channels on gpgpus.
\newblock In {\em Proceedings of the 50th annual IEEE/ACM international symposium on microarchitecture}, pages 354--366, 2017.

\bibitem{naghibijouybari2018rendered}
Hoda Naghibijouybari, Ajaya Neupane, Zhiyun Qian, and Nael Abu-Ghazaleh.
\newblock Rendered insecure: Gpu side channel attacks are practical.
\newblock In {\em Proceedings of the 2018 ACM SIGSAC conference on computer and communications security}, pages 2139--2153, 2018.

\bibitem{nai2015graphbig}
Lifeng Nai, Yinglong Xia, Ilie~G Tanase, Hyesoon Kim, and Ching-Yung Lin.
\newblock Graphbig: understanding graph computing in the context of industrial solutions.
\newblock In {\em Proceedings of the International Conference for High Performance Computing, Networking, Storage and Analysis}, pages 1--12, 2015.

\bibitem{nazaraliyev2025practical}
Ravan Nazaraliyev, Saber Ganjisaffar, Nurlan Nazaraliyev, and Nael Abu-Ghazaleh.
\newblock Practical: Subarray-level counter update and bank-level recovery isolation for efficient prac rowhammer mitigation.
\newblock {\em arXiv preprint arXiv:2507.18581}, 2025.

\bibitem{ravan2025notso}
Ravan Nazaraliyev, Yicheng Zhang, Sankha~Baran Dutta, Nael Abu-Ghazaleh, Andres Marquez, and Kevin Barker.
\newblock Not so refreshing: Attacking gpus using rfm rowhammer mitigation.
\newblock In {\em 34th USENIX Security Symposium (USENIX Security 25)}, 2025.

\bibitem{nazari2017eddie}
Alireza Nazari, Nader Sehatbakhsh, Monjur Alam, Alenka Zajic, and Milos Prvulovic.
\newblock Eddie: Em-based detection of deviations in program execution.
\newblock In {\em Proceedings of the 44th Annual International Symposium on Computer Architecture}, pages 333--346, 2017.

\bibitem{h100_cc}
{NVIDIA}.
\newblock {Confidential Compute on NVIDIA Hopper H100}.
\newblock https://images.nvidia.com/aem-dam/en-zz/Solutions/data-center/HCC-Whitepaper-v1.0.pdf.
\newblock Last accessed on: 05/25/2025.

\bibitem{cuda_programming_guide}
{NVIDIA}.
\newblock {CUDA C++ Programming Guide}.
\newblock https://docs.nvidia.com/cuda/cuda-c-programming-guide/index.html.
\newblock Last accessed on: 07/31/2024.

\bibitem{cupti}
NVIDIA.
\newblock {CUPTI}.
\newblock https://docs.nvidia.com/cupti/.
\newblock Last accessed on: 07/16/2024.

\bibitem{cuptiEventAPI}
{NVIDIA}.
\newblock {CUPTI Event API}.
\newblock \url{https://docs.nvidia.com/cupti/api/group\_\_CUPTI\_\_EVENT\_\_API.html}.
\newblock Last accessed on: 06/04/2025.

\bibitem{PTX}
NVIDIA.
\newblock {Parallel Thread Execution}.
\newblock \url{https://docs.nvidia.com/cuda/parallel-thread-execution/}.

\bibitem{cupti_doc}
NVIDIA.
\newblock {CUPTI: User Guide}.
\newblock https://docs.nvidia.com/cuda/archive/11.0\_GA/cupti/pdf/Cupti.pdf, 2022.
\newblock Last accessed on: 07/18/2024.

\bibitem{cudaSamples}
NVIDIA.
\newblock {NVIDIA CUDA samples}.
\newblock \url{https://github.com/NVIDIA/cuda-samples}, 2024.

\bibitem{nvidia_nsight_perfctr}
{NVIDIA Corporation}.
\newblock {Nsight Visual Studio Edition 4.6 User Guide: Performance Counters}.
\newblock \url{https://docs.nvidia.com/nsight-visual-studio-edition/4.6/Content/Analysis/Report/CudaExperiments/KernelLevel/PerformanceCounters.htm}, 2024.
\newblock Accessed on 05/29/2025.

\bibitem{chat_gpt}
{OpenAI}.
\newblock {Introducing ChatGPT}.
\newblock https://openai.com/index/chatgpt/.
\newblock Last accessed on: 07/31/2024.

\bibitem{ozsoy-16}
Meltem Ozsoy, Khaled~N Khasawneh, Caleb Donovick, Iakov Gorelik, Nael Abu-Ghazaleh, and Dmitry Ponomarev.
\newblock Hardware-based malware detection using low-level architectural features.
\newblock {\em IEEE Transactions on Computers}, 65(11):3332--3344, 2016.

\bibitem{mind_control_attack}
Sang-Ok Park, Ohmin Kwon, Yonggon Kim, Sang~Kil Cha, and Hyunsoo Yoon.
\newblock Mind control attack: Undermining deep learning with gpu memory exploitation.
\newblock {\em Computers \& Security}, 102:102115, 2021.
\newblock URL: \url{https://www.sciencedirect.com/science/article/pii/S0167404820303886}, \href {https://doi.org/10.1016/j.cose.2020.102115} {\path{doi:10.1016/j.cose.2020.102115}}.

\bibitem{patwari2022dnn}
Kartik Patwari, Syed~Mahbub Hafiz, Han Wang, Houman Homayoun, Zubair Shafiq, and Chen-Nee Chuah.
\newblock Dnn model architecture fingerprinting attack on cpu-gpu edge devices.
\newblock In {\em 2022 IEEE 7th European Symposium on Security and Privacy (EuroS\&P)}, pages 337--355. IEEE, 2022.

\bibitem{power_gm}
Reza Rad, Jim Plusquellic, and Mohammad Tehranipoor.
\newblock Sensitivity analysis to hardware trojans using power supply transient signals.
\newblock In {\em 2008 IEEE International Workshop on Hardware-Oriented Security and Trust}, pages 3--7, 2008.
\newblock \href {https://doi.org/10.1109/HST.2008.4559037} {\path{doi:10.1109/HST.2008.4559037}}.

\bibitem{rad2008power}
Reza~M Rad, Xiaoxiao Wang, Mohammad Tehranipoor, and Jim Plusquellic.
\newblock Power supply signal calibration techniques for improving detection resolution to hardware trojans.
\newblock In {\em 2008 IEEE/ACM International Conference on Computer-Aided Design}, pages 632--639. IEEE, 2008.

\bibitem{GPU_code_reuse_attack_2025}
Jonas Roels, Adriaan Jacobs, and Stijn Volckaert.
\newblock Cuda, woulda, shoulda: Returning exploits in a sass-y world.
\newblock In {\em Proceedings of the 18th European Workshop on Systems Security}, EuroSec'25, page 40–48, New York, NY, USA, 2025. Association for Computing Machinery.
\newblock \href {https://doi.org/10.1145/3722041.3723099} {\path{doi:10.1145/3722041.3723099}}.

\bibitem{saiz2022top}
Alvaro Saiz, Pablo Prieto, Pablo Abad, Jose~Angel Gregorio, and Valentin Puente.
\newblock {Top-down performance profiling on nvidia's GPUs}.
\newblock In {\em 2022 IEEE international parallel and distributed processing symposium (IPDPS)}, pages 179--189. IEEE, 2022.

\bibitem{salvador2007toward}
Stan Salvador and Philip Chan.
\newblock Toward accurate dynamic time warping in linear time and space.
\newblock {\em Intelligent data analysis}, 11(5):561--580, 2007.

\bibitem{sehatbakhsh2019emma}
Nader Sehatbakhsh, Alireza Nazari, Haider Khan, Alenka Zajic, and Milos Prvulovic.
\newblock Emma: Hardware/software attestation framework for embedded systems using electromagnetic signals.
\newblock In {\em Proceedings of the 52nd Annual IEEE/ACM International Symposium on Microarchitecture}, pages 983--995, 2019.

\bibitem{taneja2025roguerfm}
Hritvik Taneja and Moinuddin Qureshi.
\newblock Roguerfm: Attacking refresh management for covert-channel and denial-of-service.
\newblock {\em arXiv preprint arXiv:2501.06646}, 2025.

\bibitem{graviton}
Stavros Volos, Kapil Vaswani, and Rodrigo Bruno.
\newblock Graviton: Trusted execution environments on {GPUs}.
\newblock In {\em 13th USENIX Symposium on Operating Systems Design and Implementation (OSDI 18)}, pages 681--696, Carlsbad, CA, October 2018. USENIX Association.
\newblock URL: \url{https://www.usenix.org/conference/osdi18/presentation/volos}.

\bibitem{wang2022_pmu}
Zhendong Wang, Xiaoming Zeng, Xulong Tang, Danfeng Zhang, Xing Hu, and Yang Hu.
\newblock Demystifying arch-hints for model extraction: An attack in unified memory system, 2022.
\newblock URL: \url{https://arxiv.org/abs/2208.13720}, \href {https://arxiv.org/abs/2208.13720} {\path{arXiv:2208.13720}}.

\bibitem{wang2015simultaneous}
Zhenning Wang, Jun Yang, Rami Melhem, Bruce Childers, Youtao Zhang, and Minyi Guo.
\newblock Simultaneous multikernel: Fine-grained sharing of gpus.
\newblock {\em IEEE Computer Architecture Letters}, 15(2):113--116, 2015.

\bibitem{wei2020leaky}
Junyi Wei, Yicheng Zhang, Zhe Zhou, Zhou Li, and Mohammad~Abdullah Al~Faruque.
\newblock {Leaky DNN}: Stealing deep-learning model secret with gpu context-switching side-channel.
\newblock In {\em 2020 50th Annual IEEE/IFIP International Conference on Dependable Systems and Networks (DSN)}, pages 125--137. IEEE, 2020.

\bibitem{xu2016warped}
Qiumin Xu, Hyeran Jeon, Keunsoo Kim, Won~Woo Ro, and Murali Annavaram.
\newblock Warped-slicer: Efficient intra-sm slicing through dynamic resource partitioning for gpu multiprogramming.
\newblock {\em ACM SIGARCH Computer Architecture News}, 44(3):230--242, 2016.

\bibitem{yao2020deephammer}
Fan Yao, Adnan~Siraj Rakin, and Deliang Fan.
\newblock $\{$DeepHammer$\}$: Depleting the intelligence of deep neural networks through targeted chain of bit flips.
\newblock In {\em 29th USENIX Security Symposium (USENIX Security 20)}, pages 1463--1480, 2020.

\bibitem{dosgpu2013}
Wei Zhang.
\newblock { Defend GPUs against DoS attacks }.
\newblock In {\em 2013 IEEE 32nd International Performance Computing and Communications Conference (IPCCC)}, pages 1--2, Los Alamitos, CA, USA, December 2013. IEEE Computer Society.
\newblock URL: \url{https://doi.ieeecomputersociety.org/10.1109/PCCC.2013.6742758}, \href {https://doi.org/10.1109/PCCC.2013.6742758} {\path{doi:10.1109/PCCC.2013.6742758}}.

\bibitem{zhang2024beyond}
Yicheng Zhang, Ravan Nazaraliyev, Sankha~Baran Dutta, Nael Abu-Ghazaleh, Andres Marquez, and Kevin Barker.
\newblock Beyond the bridge: Contention-based covert and side channel attacks on multi-gpu interconnect.
\newblock In {\em 2024 International Symposium on Secure and Private Execution Environment Design (SEED)}, pages 35--36. IEEE, 2024.

\bibitem{zhang2025nvbleed}
Yicheng Zhang, Ravan Nazaraliyev, Sankha~Baran Dutta, Andres Marquez, Kevin Barker, and Nael Abu-Ghazaleh.
\newblock Nvbleed: Covert and side-channel attacks on nvidia multi-gpu interconnect.
\newblock {\em arXiv preprint arXiv:2503.17847}, 2025.

\bibitem{zhang2024invalidate}
Zhenkai Zhang, Kunbei Cai, Yanan Guo, Fan Yao, and Xing Gao.
\newblock $\{$Invalidate+ Compare$\}$: A $\{$Timer-Free$\}$$\{$GPU$\}$ cache attack primitive.
\newblock In {\em 33rd USENIX Security Symposium (USENIX Security 24)}, pages 2101--2118, 2024.

\end{thebibliography}


\end{document}